\documentclass[fleqn,usenatbib]{mnras}

\usepackage{newtxtext,newtxmath}

\usepackage[T1]{fontenc}
\usepackage{ae,aecompl}

\usepackage{graphicx}	
\usepackage{amsmath}	
\usepackage{amssymb}	

\usepackage{subcaption}
\usepackage{color}
\usepackage{nccmath}
\usepackage{float}
\usepackage{placeins}
\usepackage{arydshln}
\usepackage{pdflscape}
\usepackage{hyperref}

\title[Radio Source Extraction with \textsc{ProFound}]{Radio Source Extraction with \textsc{ProFound}}

\author[C. L. Hale et al.]{C. L. Hale,$^{1}$\thanks{E-mail: catherine.hale@physics.ox.ac.uk}
 A. S. G. Robotham,$^{2}$
 L. J. M. Davies,$^{2}$
 M. J. Jarvis,$^{1,3}$ 
\newauthor S. P. Driver$^{2}$ and I. Heywood$^{1,4}$
\\ \\
$^{1}$University of Oxford, Denys Wilkinson Building, Keble Road, Oxford, OX1 3RH, UK\\
$^{2}$International Centre for Radio Astronomy Research (ICRAR), University of Western Australia, Crawley, WA 6009, Australia \\
$^{3}$Department for Physics, University of the Western Cape, Bellville 7535, South Africa \\
$^{4}$Department of Physics and Electronics, Rhodes University, PO Box 94, Grahamstown, 6140, South Africa
}

\date{Accepted XXX. Received YYY; in original form ZZZ}

\pubyear{2018}

\begin{document}
\label{firstpage}
\pagerange{\pageref{firstpage}--\pageref{lastpage}}
\maketitle

\begin{abstract}
In the current era of radio astronomy, continuum surveys observe a multitude of objects with complex morphologies and sizes, and are not limited to observing point sources. Typical radio source extraction software generates catalogues by using Gaussian components to form a model of the emission. This may not be well suited to complicated jet structures and extended emission, particularly in the era of interferometers with a high density of short baselines, that are sensitive to extended emission. In this paper, we investigate how the optically-motivated source detection package \textsc{ProFound} \citep[][]{Robotham2018} may be used to model radio emission of both complicated and point-like radio sources. We use a combination of observations and simulations to investigate how \textsc{ProFound} compares to other source extractor packages used for radio surveys. We find that \textsc{ProFound} can accurately recover both the flux densities of simulated Gaussian sources as well as extended radio galaxies. \textsc{ProFound} can create models that trace the complicated nature of these extended galaxies, which we show is not necessarily the case with other source extraction software. Our work suggests that our knowledge of the emission from extended radio objects may be both over or under-estimated using traditional software. We suggest that {\textsc{ProFound} offers} a useful alternative to the fitting of Gaussian components for generating catalogues from current and future radio surveys. Furthermore, \textsc{ProFound}'s multi-wavelength capabilities will be useful in investigating radio sources in combination with multi-wavelength data.
\end{abstract}




\section{Introduction}
Modern radio surveys are able to combine deep and wide-area observations of the sky with greater ease than ever before. Radio facilities and the extragalactic surveys they perform such as with MeerKAT \citep[][]{Jonas2016, Jarvis2017}, Australian SKA Pathfinder \citep[ASKAP;][]{Johnston2008,Norris2011}, the Very Large Array \citep[VLA;][]{Helfand2015,Smolcic2017}, LOw Frequency ARray \citep[LOFAR;][]{LOFAR,Shimwell2017} and the Murchison Widefield Array \citep[MWA;][]{Tingay2013, Wayth2015, Hurley-Walker2017} are transforming our view of the radio skies. The increased field of view, resolution and surface brightness sensitivity of these observations allows a wide variety of complex and interesting morphologies to be observed. These include Active Galactic Nuclei (AGN) of Fanaroff-Riley Type I and II \citep[FRI and FRII;][]{Fanaroff1974} morphologies, radio relics, bent tailed radio sources as well as a large number of radio-quiet quasars and Star-Forming Galaxies (SFGs). 

With the advent of these new surveys, we are likely to observe emission that was previously unseen or unresolved, presenting more complicated morphologies than simple point sources. With this, it is important that the software used to model and generate the flux density of sources is accurate. Current software such as \textsc{PyBDSF} \citep[][]{Mohan2015} and \textsc{AEGEAN} \citep[][]{Hancock2012, Hancock2018} fit Gaussian components to radio sources to form a catalogue. For simple unresolved emission this involves fitting single Gaussian components. For resolved sources and those with extended emission and more complicated jet morphology, these are fit using a combination of Gaussian components of different sizes which are joined together to form a final source.

Whilst modelling emission with Gaussian components works well for point sources, it is not necessarily true that larger galaxies (that may appear ``disc''-like) and AGN (with extended jet morphologies) are well described through combining components. With future radio surveys, source detection algorithms using Gaussian components may struggle with more complicated structures, as well as extended emission. It is therefore important that we consider other methods to extract the flux densities of these sources. Accurate flux densities are crucial for our understanding of the extragalactic radio skies. One such need is in understanding the shape of source counts of the radio population as a whole, as well as the individual populations of radio galaxies \citep[see e.g.][]{Condon2012, Prandoni2018}. This is important in making calculations of spectral indices and modelling the spectra of sources \citep[see e.g.][]{Callingham2016, Galvin2018} and calculations of radio power and luminosity functions \citep[see e.g.][]{Mauch2007, Pracy2016,Prescott2016, Smolcic2017LF}, as well as many other investigations.

At other wavelengths, source extractors such as \textsc{SExtractor} \citep{Bertin1996} and \textsc{ProFound} \citep{Robotham2018} use pixel extraction of emission. \textsc{SExtractor} however does use ellipses to determine the total photometry whereas \textsc{ProFound} does not rely on forcing a shape to model the source and extract fluxes.  This is advantageous at optical wavelengths as galaxies have complicated structures consisting of combinations of bars, discs, spiral arms etc. The data at these wavelengths also have the advantage that the noise is less correlated and so it is easier to distinguish a galaxy detection from noise. This is more complicated for radio data where the noise is highly correlated and has Gaussian structure in it that can appear similar to real emission. This could suggest that pixel flooding detection algorithms may be less advantageous in these cases or that harsher detection criteria would be needed. This is one of the reasons that fitting Gaussian components to sources has dominated how we extract information from radio images. 

In this paper we investigate whether \textsc{ProFound} \citep[][]{Robotham2018} can be used as a source extractor for radio surveys, and the advantages it may have. \textsc{ProFound} has been used previously with optical (Turner et al. in prep) and near-IR \citep{Robotham2018,Davies2018} observations. \textsc{ProFound} will not only be especially useful for those galaxies that consist of resolved emission with more complicated shapes but also is designed with multi-wavelength galaxy studies in mind. Source information from other wavelengths can be used as a proxy for detection in another band. In the radio, for example, relationships between star formation and radio luminosity \citep{Bell2003, Garn2009, Davies2017} could be useful as proxies for radio emission. This is especially useful for future studies of galaxies, where we are ever more reliant on multi-wavelength observations. 

In this paper we first give a brief overview of the package \textsc{ProFound} and the data we use to test it in Section \ref{sec:dataandtools} before investigating how well it performs on a range of radio data. Firstly we compare how well \textsc{ProFound} performs on radio continuum imaging of the XMM-LSS field (Heywood et al. in prep) in Section \ref{sec:blindcat}. Next we investigate how well \textsc{ProFound} can recover simulated galaxies that are not limited to point sources, and include extended Gaussians, ``disc" like objects and those with jet emission, in Section \ref{sec:sims}. Finally, in Section \ref{sec:3C}, we also investigate its use on sources with known complicated jets using a handful of 3C AGN \citep[from][]{Leahy1996ATLAS}. In all these comparisons, we compare to two source detection algorithms that are widely used in radio surveys, \textsc{PyBDSF} \citep[][]{Mohan2015} and \textsc{AEGEAN} \citep{Hancock2012,Hancock2018}. We discuss the potential uses and advantages of \textsc{ProFound} in radio source extraction and draw conclusions in Section \ref{sec:discussion}.

\section{Data and Packages}
\label{sec:dataandtools}
\subsection{\textsc{ProFound}}
{\textsc{ProFound}\footnote{\url{https://github.com/asgr/ProFound}} \citep{Robotham2018} uses a method of pixel flux extraction to model galaxies, determining a ``segment'' for each object. This traces the full emission from a galaxy regardless of the shape it may have. Pixels continue to contribute to the segment until a designated flux limit above the sky is reached. 
The full details of this are given in \cite{Robotham2018} however for clarity, we present an overview of how \textsc{ProFound} creates a source catalogue here: }

\begin{enumerate}
\item {Generate a rough model of the sky through gridding up the image and using a median box-car method in order to calculate the average sky properties across the field.}
\item {Create an initial model of the sources in the image by defining those pixels which are above an assigned threshold of the rough sky model from step (i). The combined pixels which make up the source are known as a segment. Each segment includes the pixels that have started from the bright pixel which initiated the source and those pixels that have grown outwards from the bright pixel and remain above the threshold limit. }
\item {Use the source model to remove real emission and improve upon the sky model by repeating step (i).}
\item {For each source that has been defined, measure the properties.}
\item {Iterate the source finding and sky model defined in steps (i), (ii) and (iii) and dilate the segments to ensure the flux has converged to a tolerance level.}
\item {Measure the source properties of the final segments to create a final catalogue.}
\end{enumerate}

{The segment identification and dilation process \textsc{ProFound} uses to generate sources involves selecting bright pixels (above a certain sky cut) which have not been assigned to a segment yet, then searching the pixels around each segment to see if they have sufficient flux to also contribute to the source. The fact that it can grow pixels in any direction means that no morphology is assumed, this is important for extracting fluxes of complex morphological shapes. It is not limited to certain shapes and so is more naturally able to model complicated emission. This source extraction method is known to work successfully in the optical and near-IR regimes where it is easier to determine the bright emission from sources. }

{In radio images however, the Gaussian noise peaks and troughs can be misidentified as sources.} This is due to the fact that the image is convolved with the {point spread function (PSF)} which can be complicated due to the incomplete aperture. As such, source extraction software used on radio images have typically used Gaussian components with a threshold for the peak flux density per beam above a high $\sigma$ level (typically 5$\sigma$, where $\sigma$ is the rms). The pixel flux density per beam values are then extracted out to another sigma level (typically 3-4$\sigma$) and the emission is modelled as a Gaussian. These high $\sigma$ limits are used to eliminate the false detection of noise as sources. However, for bright sources with extended jet structures that we observe from radio AGN and extended emission, these are unlikely to be well represented by large Gaussian shapes. This is where the benefit of using \textsc{ProFound} may lie. 

\subsection{Radio Data}
\label{sec:data}
\subsubsection{VLA Observations of the XMM-LSS Field}
\label{sec:data_vla}
Here we make use of VLA observations of the XMM-LSS field at 1.5 GHz (Heywood et al. in prep). This covers $\sim$7.5 deg$^2$ with the central  region overlapping with the XMM-LSS field observed in the VIDEO Survey \citep{Jarvis2013}. This is a field with a wealth of ancillary data across the electromagnetic spectrum \citep[see e.g.][]{Pierre2004, Tasse2007, Mauduit2012, Davies2018, Hale2019}. The observations being used (Heywood et al. in prep) were observed with 32 pointings in B-Configuration. This reached a final rms of $\sim$16 $\mu$Jy/beam at 4.5" resolution over the VIDEO field. For our investigation, we make use of $\sim1.2 \times1.2$ deg$^2$ of this field. This was chosen so the central square degree overlaps with the CFHTLS Deep 1 Field \citep[CFHTLS D1; ][]{Cuillandre2012,Hudelot2012}, centred at (36.5$^{\circ}$, -4.5${^{\circ}}$). 

\subsubsection{Observations of 3C Sources}
\label{sec:data_3C}

To investigate how well \textsc{ProFound} can model bright, extended AGN with complex morphologies, we use observations of 3C sources \citep{Edge1959,Laing1983}. We obtained images for 5 of the 3C sources from ``An Atlas of DRAGNs" \citep{Leahy1996ATLAS}, which has information and images on 85 sources from the 3CRR sample \citep{Laing1983}. The five images used were from the first $\sim 10$ sources of the listed sources\footnote{As \textsc{ProFound} does not support the NCP projection scheme, these were 5 sources that were isolated, not in this projection scheme, had large regions of source-free sky in the cut-out and gave a variety of morphologies.}, they are: 3C16 \citep{Leahy1991}, 3C19 (see \url{http://www.jb.man.ac.uk/atlas/object/3C19.html}), 3C28 \citep[][\url{http://www.jb.man.ac.uk/atlas/object/3C28.html}]{Feretti1984}, 3C42 \citep{Leahy1991} and 3C47 \citep{Leahy1996}. We give information on the resolution and frequency on these observations in Table \ref{tab:3c_info}. Our analysis is presented in Section \ref{sec:3C}.

\begin{table}
\begin{centering}
\centering
\begin{tabular}{c c c c }
Source & Reference & Resolution & Frequency  \\ 
& & (") & (MHz) \\ \hline \hline
3C16 &\cite{Leahy1991}&1.25 &1477 \\
3C19 &see \url{http://www.jb.man.ac.uk/}&0.15&1534 \\
& \url{atlas/object/3C19.html} & &  \\
3C28 &\cite{Feretti1984}&1.10&1424 \\
& \url{http://www.jb.man.ac.uk/} && \\
& \url{atlas/object/3C28.html} & &  \\
3C42 &\cite{Leahy1991}&1.20 &1477 \\
3C47 &\cite{Leahy1996}&1.00 &1650 \\
\end{tabular}
\end{centering}
\caption{Information on the 3C observations that have been used in Section \ref{sec:3C}. For each source the resolution, frequency and reference are given \citep[from][]{Leahy1996ATLAS}.}
\label{tab:3c_info}
\end{table}

\section{Source Detection Parameters}
\label{sec:detect_param}

In order to compare \textsc{ProFound} to other source extractors, it is necessary to determine which parameters to use in order to make comparisons. We use two approaches to do this. 

The first is to determine the \texttt{skycut} parameter which is necessary for \textsc{ProFound} to be used and then compare these to the typical default parameters that are used for \textsc{PyBDSF} and \textsc{AEGEAN}. Both \textsc{PyBDSF} and \textsc{AEGEAN} have been used in past radio continuum observations \citep[see e.g.][]{Hurley-Walker2017,Shimwell2017} and have been compared to each other in previous tests for large survey data challenges \citep[a comparison for use on simulated images in preparation for EMU was performed in][]{Hopkins2015}. This gives default parameters for \textsc{PyBDSF} of \texttt{thresh\_isl=3.0} and \texttt{thresh\_pix=5.0}. \textsc{AEGEAN} used the parameters: \texttt{floodclip}=4.0 and \texttt{seedclip}=5.0. 

The second is to compare how well the source extraction algorithms model sources when using parameters that have similar rates of false detections. This is to attempt to tailor the software to the VLA image in order to provide a more fair comparison to one another, where they have similar accuracy detecting real emission. We describe how we determined these parameters below. 

\begin{figure*}	
\begin{center}
\begin{minipage}[b]{0.5\textwidth}
\centering
\includegraphics[width=8cm]{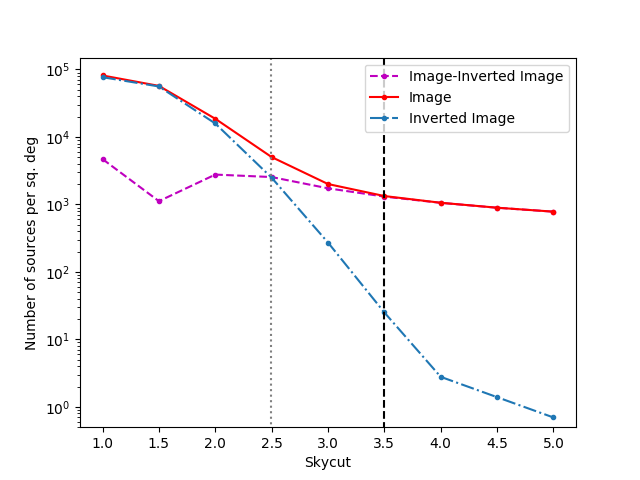}
\subcaption{}
\end{minipage}%
\begin{minipage}[b]{0.5\textwidth}
\centering
\includegraphics[width=8cm]{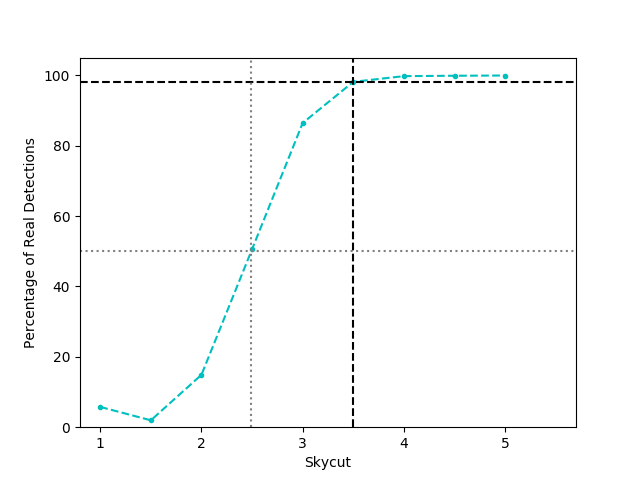}
\subcaption{}
\end{minipage}%
\caption{{The false detection rate of sources with \textsc{ProFound} when sources are extracted from the VLA image of the XMM-LSS field (Heywood et al. in prep), described in Section \ref{sec:data_vla}}. Shown in (a) is the number density of sources in the image (red) and the inverse image (blue) as a function of the \textsc{ProFound} parameter \texttt{skycut}, as well as the difference between the red and blue lines (magenta); (b) the percentage of false detections as a function of \texttt{skycut}. For clarity, the grey dotted lines indicates where a 50\% percentage of real detections occur and the black dashed line indicates the \texttt{skycut} of 3.5, chosen for this investigation, at a false detection rate of $\sim$2\%.}
\label{fig:fdr_vs_skycut}
\end{center}
\end{figure*}

\begin{figure*}	
\begin{center}
\centering
\includegraphics[width=16cm]{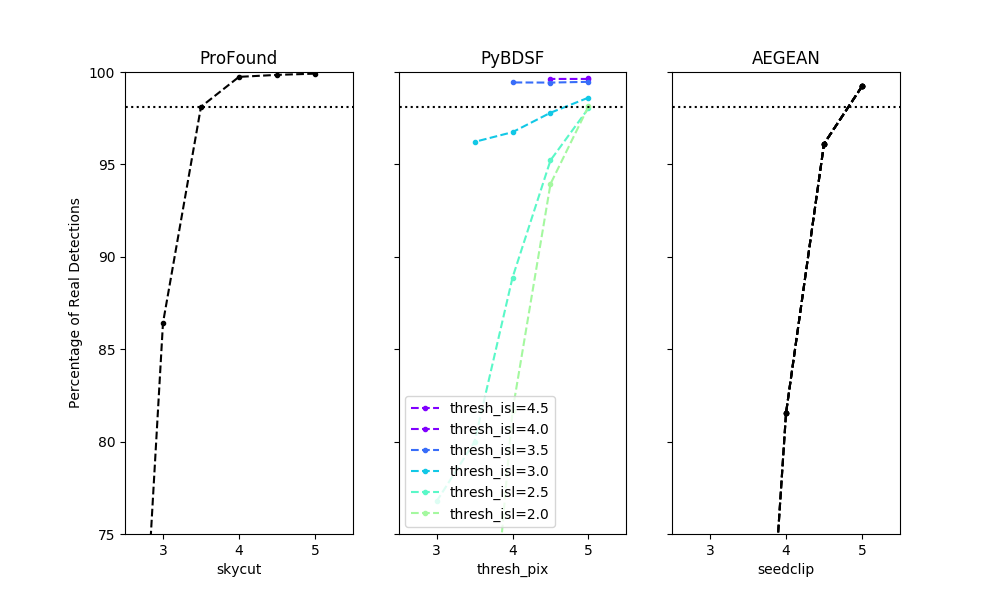}
\caption{The percentage of real detections for the different source extraction software used when their respective detection parameters are varied. This is shown for \textsc{ProFound} (left), \textsc{PyBDSF} (centre) and \textsc{AEGEAN} (right).}
\label{fig:fdr_source_extractor}
\end{center}
\end{figure*}

\subsection{\textsc{ProFound}}
\label{sec:detect_param_profound}
To decide on the necessary detection parameters for \textsc{ProFound} we consider the false detection of sources. As mentioned previously, the correlated noise in radio data means that it is likely a threshold larger than used in optical and IR surveys will be necessary. We need to ensure that these sky cuts are not too extreme such that we are unable to flood the pixels to extract emission across the source. {\texttt{skycut} here is the number of $\sigma$ (the sky rms) to be included in the source.} In order to determine which \texttt{skycut} is appropriate with these observations, we investigate how our false detection varies for different \texttt{skycut} values. This allows us to determine at which point \textsc{ProFound} becomes limited by the correlated noise of the image and is picking up too many noise spikes as sources. \texttt{skycut} is the parameter in \textsc{ProFound} which determines how many $\sigma$ above the sky a pixel can contribute to the source segment. Varying this will determine both the number of sources as well as how many pixels are combined together to extract the total flux density of the source. For bright objects, the majority of this emission will be significantly above the sky and so \textsc{ProFound} will measure its total flux density.

To quantify the false detection rate, we use the assumption that the noise in the image consists of Gaussian peaks and troughs that are symmetric. This symmetry means that a negative version of the image (from now on known as inverted) has the same noise properties as the non-inverted image, meaning large noise troughs in the original image are now detectable as sources. Hence, by running \textsc{ProFound} (and the other software) on the inverted image, the number of detected sources should be approximately equal to the number arising from false positive noise spikes in the true image. 

Using this, it is possible to constrain the percentage of ``real" detections in the image, as in Equation \ref{eq:real}. By investigating how the false detection varies with \texttt{skycut} (in steps of 0.5) this can allow us to pick an optimum value of \texttt{skycut} that successfully extracts sources with {minimal} contamination from noise. 

\begin{equation}
\% \ Real \ Detections = 100 \times \frac{N_{image} - N_{inv. \ image}}{N_{image}}
\label{eq:real}
\end{equation}

The results from investigating the percentage of real detections with \textsc{ProFound} can be seen in Figure \ref{fig:fdr_vs_skycut}, where the left hand panel (Figure \ref{fig:fdr_vs_skycut}a) shows the number of sources detected per square degree. This is shown for both the image (red) as well as the inverted image (blue). The difference between the numbers detected in the original and inverted images is also shown (magenta). The right hand panel (Figure \ref{fig:fdr_vs_skycut}b) shows the percentage of real detections (as described in Equation \ref{eq:real}) compared to the \texttt{skycut}.

As expected, the percentage of real detections is near 100\% for high \texttt{skycut} values, where a conservative $\sigma$ cut is used. The number of sources in the image that we believe to be real declines sharply below a \texttt{skycut} of 3.5. At a \texttt{skycut} of 3.5, we have a $\sim 98\%$ real detection percentage in our catalogues, this declines to $\sim 87\%$ at a \texttt{skycut}=3 and to 50\% and below for \texttt{skycut} $<$2.5. As such a \texttt{skycut} of 3.5 is appropriate to use with \textsc{ProFound} on the data in order to minimize the number of false detections. This is used in all future work unless otherwise stated\footnote{{We note that this value of \texttt{skycut} was appropriate here, but may not be for data which has more contamination from e.g. sidelobes in the image.}}.

We therefore use the following prescription in running \textsc{ProFound} on the radio images:
\begin{enumerate}
\item Run a blind detection with \textsc{ProFound} using \texttt{profoundProFound}. Run this over the image using \texttt{skycut=3.5} and set \texttt{groupstats=TRUE} and \texttt{groupby=`segim'}. Each source is defined as a segment. 
\item {Using the grouped segments (\texttt{group\$groupim}) and the corresponding statistics (properties) for these segments from \texttt{groupstats}, a catalogue of sources can be defined. Using the grouped segments ensures that any adjacent segments are combined and the source information for the merged segment is recorded within the source catalogue. }
\item Apply a beam correction to convert between the map (in Jy/beam) to the total flux densities (in Jy).
\end{enumerate}

Step (ii) ensures that neighbouring segments are combined together. Having used \texttt{groupstats=TRUE} and \texttt{groupby=`segim'}  allows a segmentation map (a map of the segments) to be generated in which all segments that are touching are joined together into one single object. As segments are determined by bright emission, many locations within a single object could be defined as a separate segment. Due to the on-sky density of radio sources at these flux densities, we are unlikely to have emission that is adjacent but not from the same source. {However, where data is confused due to the resolution and sensitivity of the observations, combining segments together may not be appropriate.} Step (iii) corrects the flux densities in the extracted \textsc{ProFound} catalogue from Jy/beam to Jy. This is a simple numerical conversion which is applied after the source catalogue has been generated.

For steps (i) and (ii) we present the commands used in \textsc{ProFound} to obtain the extracted catalogue, for clarity:
\begin{itemize}
\footnotesize
\item \texttt{image=readFITS(image\_file)}
\item \texttt{image\_blind=profoundProFound(image, plot=FALSE, skycut=3.5, rotstats=TRUE, boundstats=TRUE, nearstats=TRUE, groupstats=TRUE, groupby=`segim', verbose=TRUE)}
\item \texttt{write.csv(image\_blind\$groupstats, file=`file\_name.csv', quote=FALSE, row.names=FALSE)}
\end{itemize} \normalsize
To make the source model: 
\begin{itemize} \footnotesize
\item \texttt{segim\_model=image\_blind\$group\$groupim} 
\item \texttt{segim\_model[image\_blind\$group\$groupim!=0]=1} 
\item \texttt{segim\_model[image\_blind\$group\$groupim==0]= \linebreak \indent \indent \indent as.numeric(NaN)} 
\item \texttt{model=(image\$imDat-image\_blind\$sky)*segim\_model} 
\normalsize
\end{itemize}

{Again, we note that in both existing and future observations where the radio data is confused, using \texttt{groupstats=TRUE} and \texttt{groupby=`segim'} may not be appropriate, as the source density may be too high. }

\subsection{\textsc{PYBDSF} and \textsc{AEGEAN}}
\label{sec:detect_param_pybdsf}
To make appropriate comparisons between the source detection packages we also calculate the real detection fraction for \textsc{PyBDSF} and \textsc{AEGEAN}. This is to choose parameters in both \textsc{PyBDSF} and \textsc{AEGEAN} that give similar percentages of real detections to \textsc{ProFound}.
 
\subsubsection{\textsc{PyBDSF}}
For \textsc{PyBDSF} we only change the parameters \texttt{thresh\_isl} and \texttt{thresh\_pix}. Of these parameters, \texttt{thresh\_isl} determines the number of sigma that the boundary of the source can flood out to for the pixels to be included in the fitting. On the other hand, \texttt{thresh\_pix} helps to determine whether a source is included in the catalogue. \textsc{PyBDSF} uses an absolute thresholding to quantify whether a source is determined to be detected. This absolute thresholding only includes sources in the final catalogue with fluxes $>$ \texttt{thresh\_pix $\times$ rms +mean(map)}. As the mean map value within an island is smaller if \texttt{thresh\_isl} is smaller, due to more lower flux pixels in the source, more sources will be detected for the same \texttt{thresh\_pix} but with smaller \texttt{thresh\_isl}. Therefore for \textsc{PyBDSF} both \texttt{thresh\_isl} and \texttt{thresh\_pix} will affect the number of false detections. Although other parameters can be changed, using the default settings and only varying the threshold limits should give a good comparison between the source extractors, as the complexities of varying all the parameters can be a long process and so most users are likely to only change a handful of parameters. For \textsc{PyBDSF} we output each source catalogue for the different thresholds, where overlapping Gaussian components that \textsc{PyBDSF} has designated to be part of the same source have been combined together\footnote{see \url{http://www.astron.nl/citt/pybdsm/algorithms.html\#grouping-of-gaussians-into-sources} for how \textsc{PyBDSF} groups sources.}. 

As in Section \ref{sec:detect_param_profound}, we construct the percentage of real detections using \textsc{PyBDSF} but now as a function of \texttt{thresh\_isl} and \texttt{thresh\_pix} (again in steps of 0.5). This can be seen in Figure \ref{fig:fdr_source_extractor} (middle panel). For \textsc{PyBDSF} two parameters are varied and the percentage of real detections for given \texttt{thresh\_pix} values, with varying \texttt{thresh\_isl} are shown in different colours ranging from a value of 2 (light green) to 4.5 (purple) for \textsc{PyBDSF}. The black dashed horizontal line in all three panels indicates the percentage of real detections for the value of \texttt{skycut} that we use for our \textsc{ProFound} detections. We plot these only for values where \texttt{thresh\_pix}>\texttt{thresh\_isl}. 

As can be seen in Figure \ref{fig:fdr_source_extractor}, to obtain similar percentages of real detections for \textsc{PyBDSF} then either values of \texttt{thresh\_isl}/\texttt{thresh\_pix} of 2.5/5.0 or 3.0/4.5 should be used. The first of these combinations gives a percentage of real detections most similar to that obtained with \textsc{ProFound}, however we choose to use the 3.0/4.5 combination which has a lower $\sigma$ threshold. This means that more sources will be detected, in this case N$_{3.0/4.5} \sim$ 1.1 $\times$ N$_{2.5/5.0}$ more sources. 

\subsubsection{Extended emission with \textsc{PyBDSF}}
\label{sec:extended_pybdsf}
We also consider using settings which allow \textsc{PyBDSF} to better model extended structures in the image (for further information see \url{http://www.astron.nl/citt/pybdsm/examples.html#image-with-extended-emission}). To do this, we run \textsc{PyBDSF} with the settings described previously but with also \texttt{flagging\_opts=True}, {\texttt{flag\_maxsize\_bm=100}}, \texttt{atrous\_do=True}, \texttt{rms\_map=False}, \texttt{mean\_map=`zero'}. We will refer to all tests using this as \texttt{atrous\_do} from now. \texttt{flag\_maxsize\_bm=100} allows for large Gaussians, much greater than the beam size, to be fit whilst \texttt{atrous\_do=True} allows Gaussians of different scales to be fit. Setting \texttt{mean\_map=`zero'} ensures the background mean is set to 0, which is helpful if there is extended emission that could be misinterpreted as background. 

\subsubsection{\textsc{AEGEAN}}
For \textsc{AEGEAN} there are again two main parameters that we consider changing, similar to \textsc{PyBDSF} these are \texttt{floodclip} and \texttt{seedclip}. \texttt{floodclip} is similar to \texttt{thresh\_isl} and \texttt{seedclip} is similar to \texttt{thresh\_pix} as used in \textsc{PyBDSF}. We again calculate the percentage of real detections, but this time as a function of \texttt{seedclip} only (again in steps of 0.5). \textsc{AEGEAN} has a fixed thresholding based solely on the \texttt{seedclip} value and so whilst \texttt{floodclip} will determine the extent to fit sources to, it will not affect the number of sources detected. This shown in Figure \ref{fig:fdr_source_extractor} (right hand panel). A value of \texttt{seedclip} between 4.5 and 5.0 seems appropriate for \textsc{AEGEAN}. As this does not depend on \texttt{floodclip}, we use the default value of 4.0 for this. Although a seedclip of 4.5 has a slightly smaller real percentage fraction of detections, this is still a high value $\sim 96\%$. As a value of \texttt{seedclip}=5.0 is the default value, we shall use here a value of \texttt{seedclip}=4.5 as a comparison.  \\ \\

\noindent Now that the different parameters for the different source detection software have been determined, we will use these parameters, unless otherwise stated. We shall compare in all cases both using the default parameters as well as the parameters from the real source detection analysis\footnote{{We note that when comparing the false detections over the central square degree only yielded the same detection parameters choice as determined here.}}. 

\begin{figure*}
\begin{center}

\begin{minipage}[b]{\textwidth}
\centering
\includegraphics[height=3.2cm]{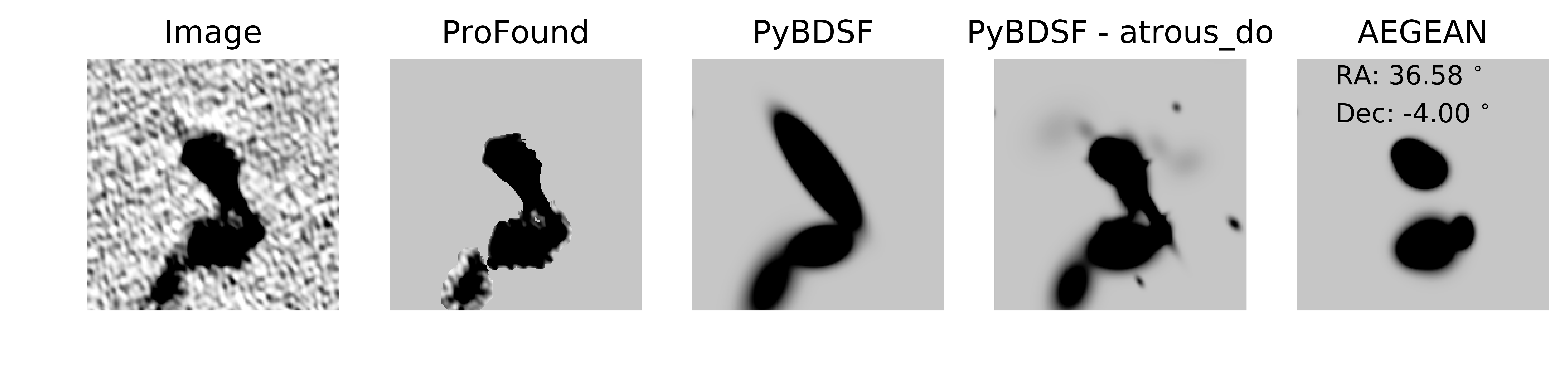}
\end{minipage} \newline
\begin{minipage}[b]{\textwidth}
\centering
\includegraphics[height=3.2cm]{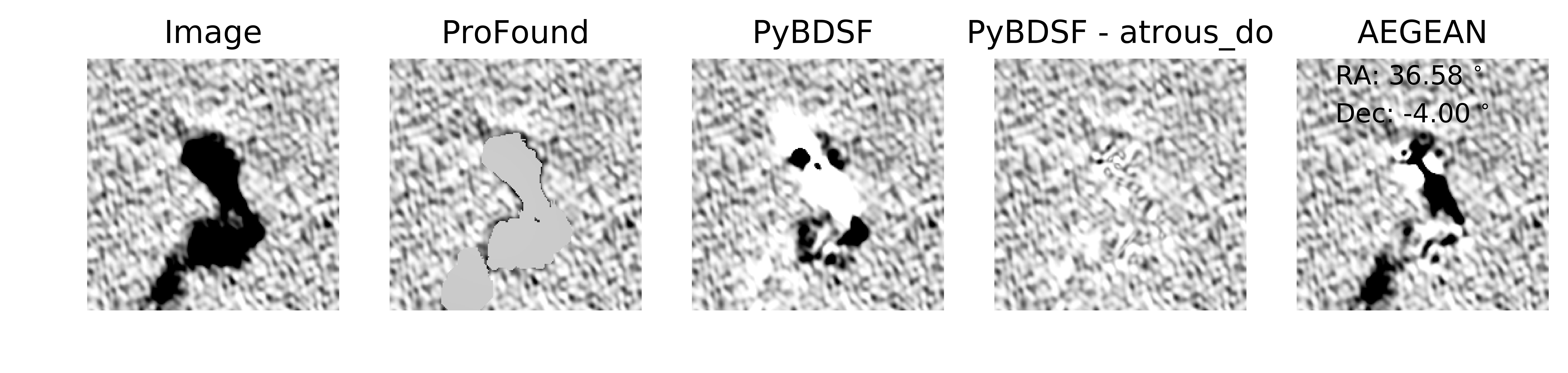}
\subcaption{}
\end{minipage}
\newline \newline
\begin{minipage}[b]{\textwidth}
\centering
\includegraphics[height=3.2cm]{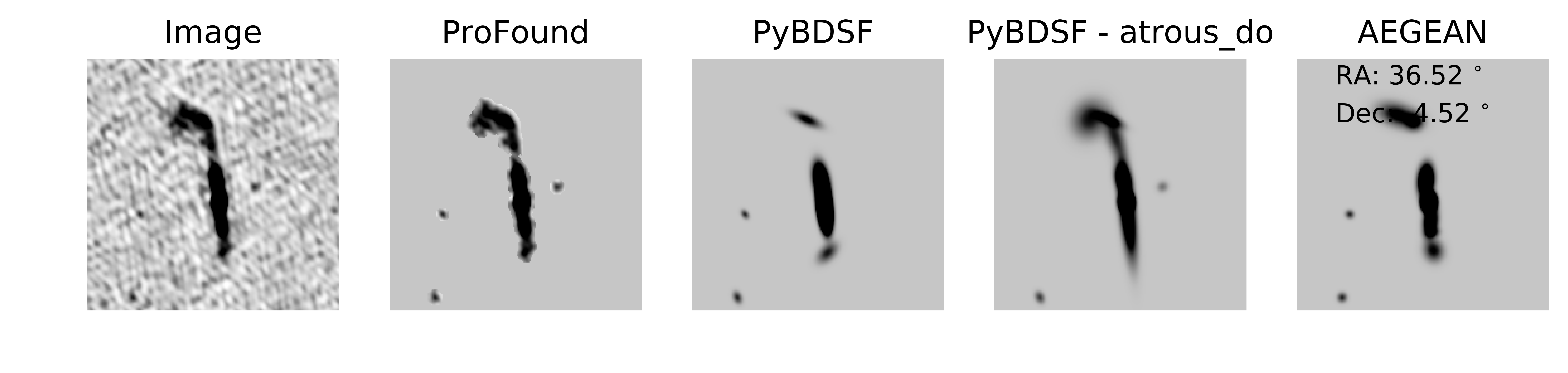}
\end{minipage} \newline
\begin{minipage}[b]{\textwidth}
\centering
\includegraphics[height=3.2cm]{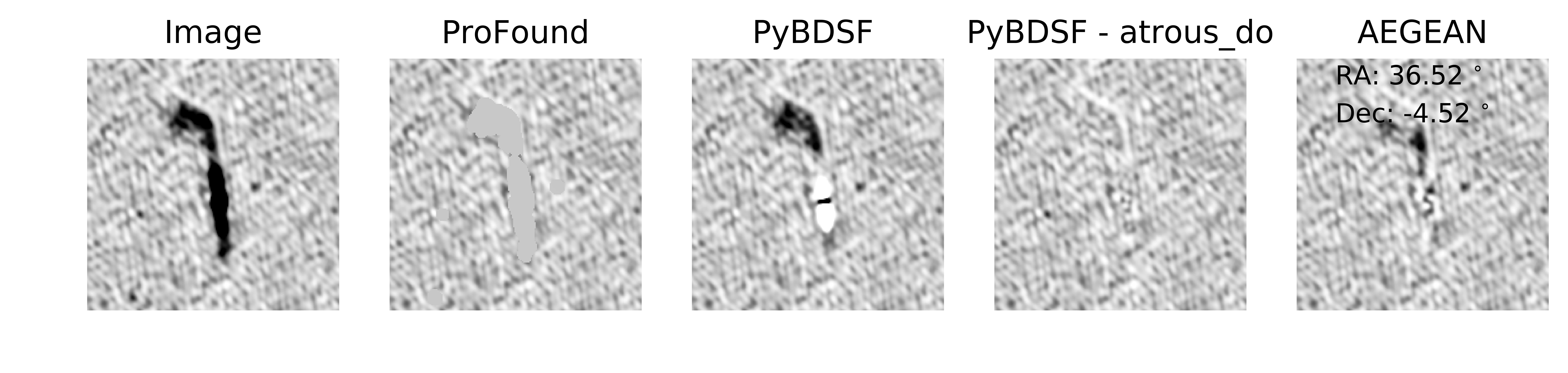}
\subcaption{}
\end{minipage}
\newline \newline
\begin{minipage}[b]{\textwidth}
\centering
\includegraphics[height=3.2cm]{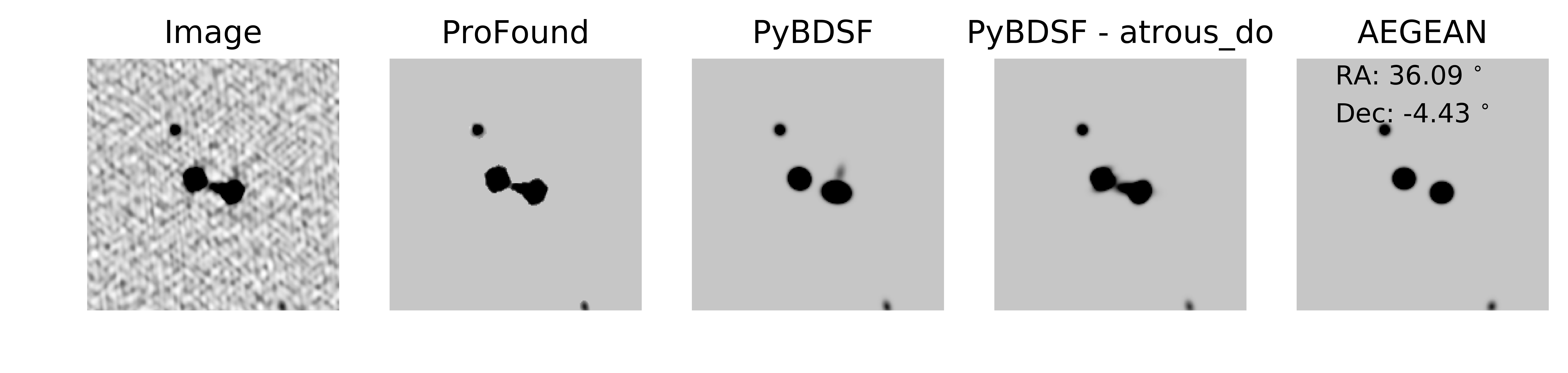}
\end{minipage} \newline
\begin{minipage}[b]{\textwidth}
\centering
\includegraphics[height=3.2cm]{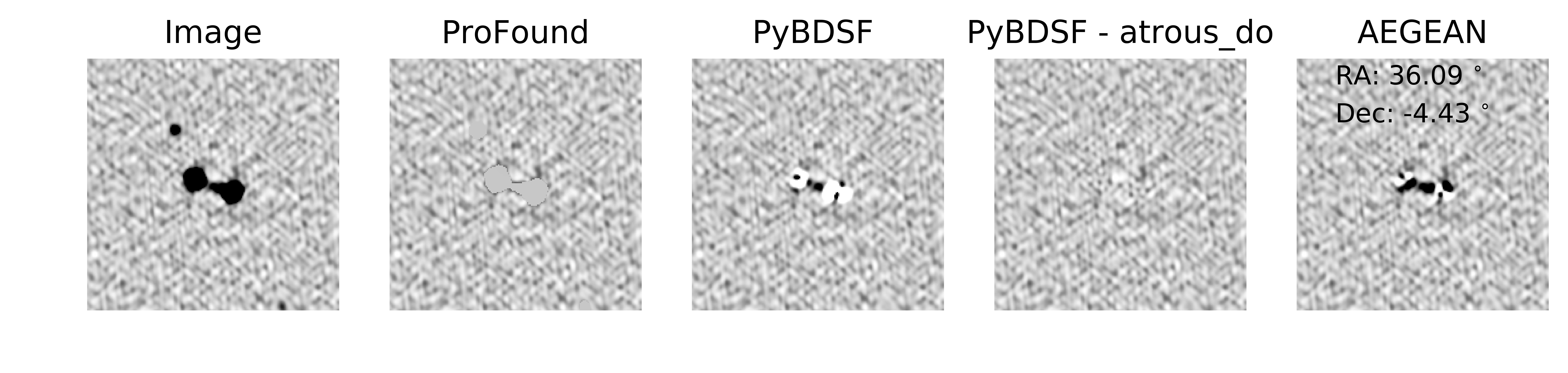}
\subcaption{}
\end{minipage}
\caption{Example images of three extended morphology radio sources in the data (left panel) and their corresponding modelled emission in \textsc{ProFound} (centre left - \texttt{skycut=3.5}), \textsc{PyBDSF} (centre - without \texttt{atrous\do}; centre right - using \texttt{atrous\_do}) and \textsc{AEGEAN} (right) from the blind detections. For \textsc{PyBDSF} and \textsc{AEGEAN} the default parameters were used. These images were selected to highlight where \textsc{ProFound} may have an advantage in source detection compared to source extractors where Gaussian components are joined together to form a source. The flux density per beam scale for each of the images is the same and ranges between -0.05 mJy/beam and 0.1 mJy/beam.}
\label{fig:sources}
\end{center}
\end{figure*}

\begin{figure*}	
\begin{center}
\begin{minipage}[b]{\textwidth}
\centering
\includegraphics[width=14cm]{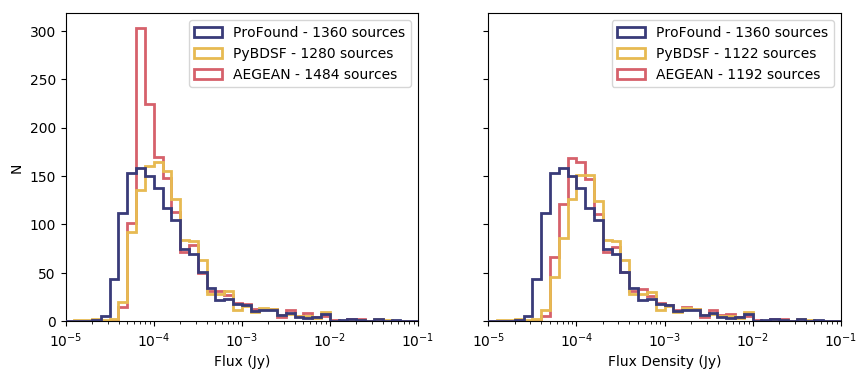}
\subcaption{Without \texttt{atrous\_do} used for \textsc{PyBDSF}.}
\end{minipage}%
\newline
\begin{minipage}[b]{\textwidth}
\centering
\includegraphics[width=14cm]{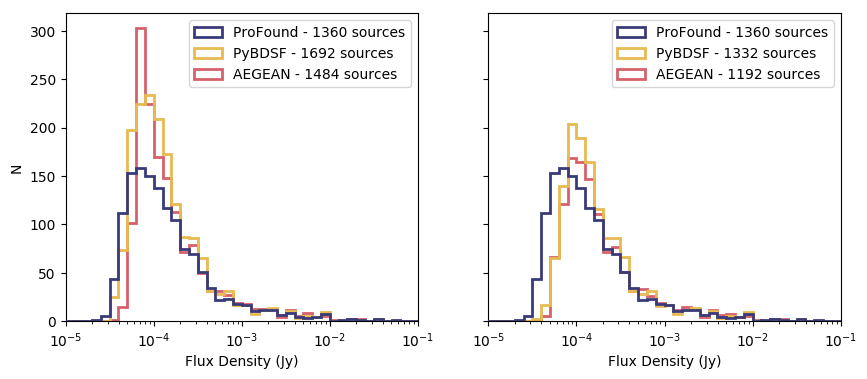}
\subcaption{With \texttt{atrous\_do} used for \textsc{PyBDSF}.}
\end{minipage}%
\caption{Histogram of the flux densities for the sources recovered in the source catalogues from \textsc{ProFound} (blue), \textsc{PyBDSF} (yellow) and \textsc{AEGEAN} (red). The number of objects in the catalogue from each source extractor is shown in the top left hand corner of each panel. On the left hand panel the results when the parameters determined in Section \ref{sec:detect_param_pybdsf} are used and the right hand panel shows this with the default parameters of \textsc{PyBDSF} and \textsc{AEGEAN} used to generate the source models. For the top panel (a) \textsc{PyBDSF} has the \texttt{atrous\_do} setting off whilst it is used in the bottom panel (b).}
\label{fig:hist_fluxes}
\end{center}
\end{figure*}

\begin{figure*}	
\begin{center}
\begin{minipage}[b]{\textwidth}
\centering
\includegraphics[width=13cm]{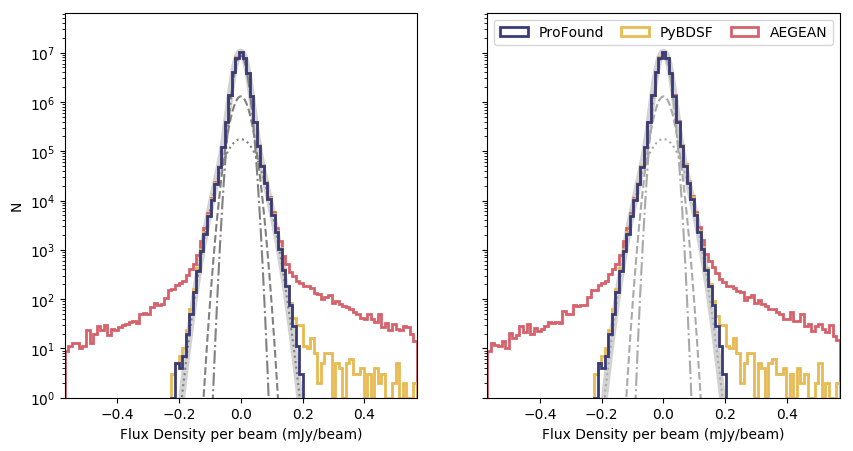}
\subcaption{Without \texttt{atrous\_do} used for \textsc{PyBDSF}.}
\end{minipage}%
\newline
\begin{minipage}[b]{\textwidth}
\centering
\includegraphics[width=13cm]{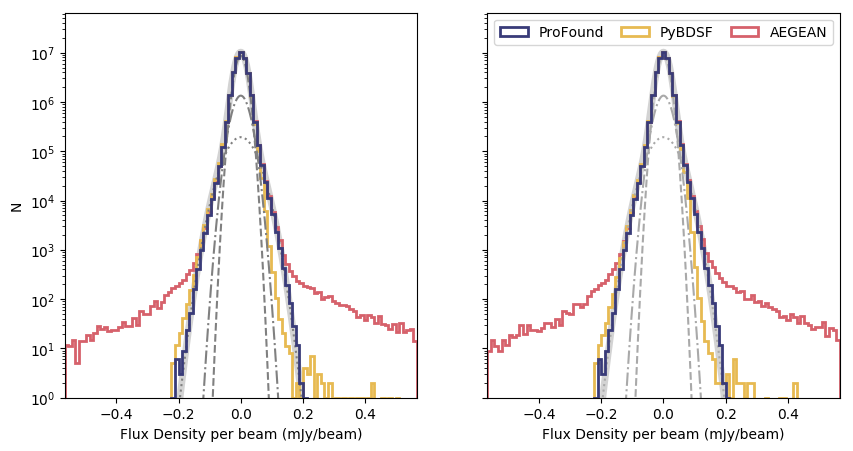}
\subcaption{{With \texttt{atrous\_do} used for \textsc{PyBDSF}.}}
\end{minipage}%
\caption{Histogram of the residual images created by \textsc{ProFound} (blue), \textsc{PyBDSF} (yellow) and \textsc{AEGEAN} (red). A model of the noise in the image is also shown through a three Gaussian model (grey dotted, dashed and dot-dashed lines) with the sum of these three components shown in the thicker grey solid line. The left hand panel shows the results when the parameters determined in Section \ref{sec:detect_param_pybdsf} are used for source extraction and the right hand panel shows this with the default parameters of \textsc{PyBDSF} and \textsc{AEGEAN} used to generate the source models. The top panel shows the residuals when \texttt{atrous\_do} is not used in \textsc{PyBDSF}, whilst the bottom panel shows the results when the \texttt{atrous\_do} setting is switched on. }
\label{fig:hist_residuals}
\end{center}
\end{figure*}

\section{Blind Detection Using the Different Source Extractors}
\label{sec:blindcat}

When \textsc{ProFound} is run over the XMM-LSS image, 1,360 sources were found over the central 1 deg$^2$ of this field\footnote{That overlaps with the CFHTLS Deep 1 field.}. For the same region \textsc{PyBDSF} found 1,122 sources (1,332 using \texttt{atrous\_do}) and 1,192 with \textsc{AEGEAN}, when the default parameters were used. If we instead compare the catalogues for the parameters based on the percentage of real detections, \textsc{PyBDSF} found 1,280 sources  (1,692 using \texttt{atrous\_do}) where as \textsc{AEGEAN} found 1,484. Differences in these numbers will arise from differences in the detection depths of the different algorithms as well as differences in whether resolved sources have been split into multiple components. As we feel that the benefits of \textsc{ProFound} may arise from its ability to determine flux densities and models of sources with complicated morphology and extended emission, we include images in Figure \ref{fig:sources} for three of these extended sources. This is to compare the models from \textsc{ProFound} to \textsc{PyBDSF} and \textsc{AEGEAN}. These are shown when the default parameters of \textsc{PyBDSF} and \textsc{AEGEAN} and a \texttt{skycut} value of 3.5 for \textsc{ProFound} were used.

As can be seen from the examples in Figure \ref{fig:sources}, \textsc{ProFound}, in these cases, captures the shape of these sources that have complicated morphologies. With \textsc{AEGEAN} and \textsc{PyBDSF} (without \texttt{atrous\_do}) in the cases shown, parts of the source are not well modelled and do not capture the full shapes. This may be due to over-fitting of components, such as in Figure \ref{fig:sources} (a) with \textsc{PyBDSF} or due to under fitting of components, as in Figure \ref{fig:sources}(b). With \textsc{PyBDSF} when \texttt{atrous\_do} is used, these sources are much better modelled and the extended emission is better captured, as seen in Figures \ref{fig:sources}(b) and (c). However, it is noticeable in \ref{fig:sources}(a), that there can be thin, extended haloes around these sources due to some of the Gaussian components being fit. This is not the case with \textsc{ProFound}. The residuals for \textsc{ProFound} {(i.e. the sky background image (over the segments) as the sources have been subtracted out)} are smooth and close to zero and do not show the noise structure that is shown in the residuals from \textsc{PyBDSF} and \textsc{AEGEAN}. This is because the sky is modelled as a smooth distribution and is subtracted from the images. This means that in the case of \textsc{ProFound}, noise sub-structure is likely to be contained within the model. However if the noise is symmetric over the source, it should approximately sum to zero and as such not affect the estimate of the total flux density, {although this noise may affect the measured peak flux}. 

To compare how each software has extracted sources quantitatively, we consider both the flux densities in the catalogues as well as the residuals of the images once the sources have been extracted. In terms of the total flux densities, we present a histogram of these from the different source extraction software in Figure \ref{fig:hist_fluxes}. This is again shown for the parameters determined in Section \ref{sec:detect_param_pybdsf} (left) and for the default parameters (right). At high flux densities, where we expect that all three algorithms should easily detect sources, the histograms are similar. The differences occur at lower flux densities where, in both cases, \textsc{ProFound} appears to preferentially detect fainter sources compared to the other two detection algorithms. This is likely due to the lower \texttt{skycut} that allows a source to be classified i.e. using 3.5 compared to the 4.5/5$\sigma$ peak detection threshold with \textsc{PyBDSF} and \textsc{AEGEAN}. However, at $\sim 10^{-4}$Jy, \textsc{ProFound} appears to be finding significantly fewer sources. This could suggest that \textsc{ProFound} is being limited by not being able to probe the full emission of a source over the beam area (if it is at low signal to noise), which may explain the large number of sources with faint flux densities. If this is the case and \textsc{ProFound} is unable to fully sample the full beam, we can correct for this. This is discussed in Section \ref{sec:beam_correction}.

The histogram of the residuals {(across the entire image, not just the central 1deg$^2$)} from each model is also shown in Figure \ref{fig:hist_residuals}. If all sources in the image have been extracted successfully then the residuals (image - model) should follow a Gaussian distribution. Any deviation from this suggests either an under or over-fitting of sources. In Figure \ref{fig:hist_residuals} we show the results using the parameters from Section \ref{sec:detect_param_pybdsf} (left) default parameters (right) and also show the results of using \textsc{PyBDSF} without (top) and with (bottom) \texttt{atrous\_do} turned on. {We also include, in Figure \ref{fig:hist_residuals}, a model for Gaussian noise in the image by fitting the negative residuals (as these have less of an excess tail) from \textsc{PyBDSF} as a Gaussian of variable amplitude and $\sigma$. As the region of VLA image that we use for this work is noisier at higher declinations due to primary beam corrections at the edge of the mosaiced pointings, we do not expect it to be perfectly modelled as a Gaussian. Due to this, the Gaussian noise is modelled as a combination of multiple (three) Gaussian components which can be seen by the dashed, dotted and dot-dashed lines. The combined noise model is shown in the thick grey line. The lowest noise component fit here has a noise value of $\sim 0.016$ mJy/beam, with the other components having noise levels of $\sim 0.022$ mJy/beam and $\sim 0.040$ mJy/beam. }

From Figure \ref{fig:hist_residuals}(a) it appears that the residuals from \textsc{ProFound} and also \textsc{PyBDSF} (with \texttt{atrous\_do} turned on) are much more similar to a symmetric multi-Gaussian distribution than for \textsc{PyBDSF} (without \texttt{atrous\_do}, Figure \ref{fig:hist_residuals}b) and \textsc{AEGEAN}. \textsc{ProFound} and \textsc{PyBDSF} (with \texttt{atrous\_do}) do not show the excess of positive residuals that both \textsc{PyBDSF} and \textsc{AEGEAN} show. This suggests that \textsc{ProFound} is able to successfully model sources in this field, and leaving only small residuals. The large number of positive residuals that remain from \textsc{AEGEAN} and \textsc{PyBDSF} (without \texttt{atrous\_do}) suggest the models are under-fitting the sources in the field. With \texttt{atrous\_do} switched on, \textsc{PyBDSF} has less excess positive residuals compared to \textsc{ProFound}, but slightly more negative residuals. This suggests that the residuals are not symmetric and may suggest an over fitting of sources with \textsc{PyBDSF}. 

\subsection{Beam Correction}
\label{sec:beam_correction}
As mentioned in Section \ref{sec:blindcat}, if \textsc{ProFound} is not able to fully explore the full beam of a faint, unresolved source it may underestimate the source flux density. {This is because the \texttt{skycut} level will be a larger fraction of the peak flux for these sources and so the flux in the wings of the source are unlikely to be included}. Fortunately, this can be easily accounted for. Given knowledge of the beam shape, we create a model Gaussian for the beam. Using the source segmentation mask, and centering the Gaussian beam on the \texttt{RAcen} and \texttt{Deccen} position, from the \textsc{ProFound} catalogue, {for each source in the catalogue (regardless of shape) we then calculate what fraction of the beam flux is observed within the segment. This correction factor will be negligible for bright sources and for extended sources but will be larger for the fainter, unresolved sources.}

\begin{figure*}	
\begin{center}
\centering
\includegraphics[width=18cm]{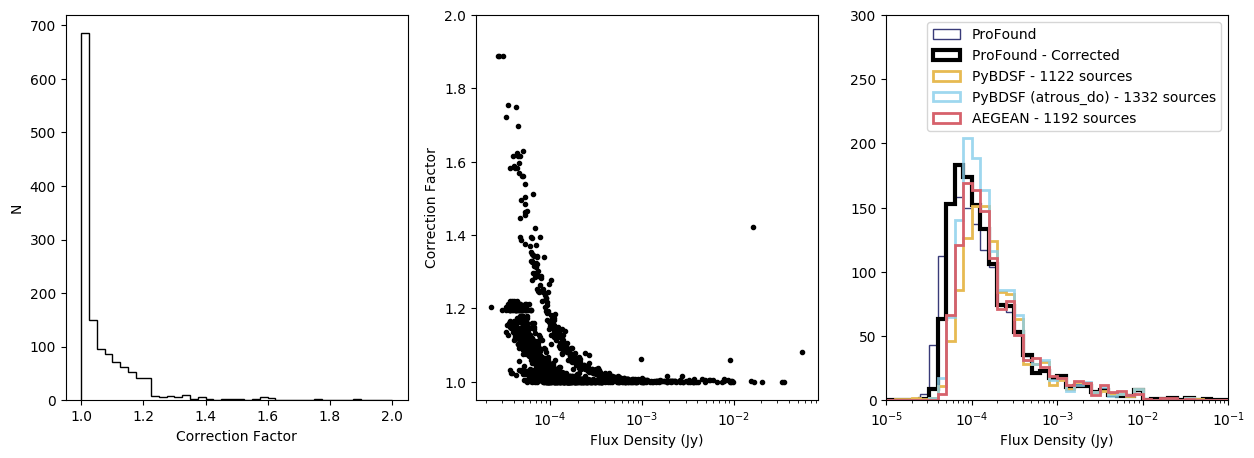}
\caption{The correction factor applied to the blind detection to ensure they have fully sampled the beam. On the left hand panel is a histogram of the correction factors that need to be applied to each source. The centre panel shows this correction factor as a function of flux density. The right hand panel shows the source counts with the corrected source counts from \textsc{ProFound} (black), \textsc{ProFound} with no correction applied (thin, blue), \textsc{PyBDSF} (without \texttt{atrous\_do}, yellow), \textsc{PyBDSF} (with \texttt{atrous\_do}, light blue)  and \textsc{AEGEAN} (red), when the default parameters are used.}
\label{fig:beamcorr}
\end{center}
\end{figure*}

A histogram of the corresponding correction factors generated for this blind catalogue {(within the central 1deg$^2$)} can be seen in Figure \ref{fig:beamcorr} (left), this correction factor is also shown as a function of the uncorrected flux density in Figure \ref{fig:beamcorr} (centre). This central panel shows there are multiple tracks in the correction factors as a function of flux. These are thought to arise from the fact that there are discrete pixels included in the segments and this will impact the fraction of the beam included in the source, depending on the noise levels at the source location. The majority of correction factors are $\sim1$, corresponding to about half of the sources, however there are significant numbers of correction factors up to $\sim 1.2$. These correction factors are typically higher for the fainter sources, that are more likely to have pixel values closer to the noise limit. We apply the correction factors to our flux densities and re-plot the flux histograms from Figure \ref{fig:hist_fluxes} in Figure \ref{fig:beamcorr} (right), the corrected flux densities are now shown in black. The flux densities from \textsc{ProFound} are now more similar to those from \textsc{PyBDSF} and \textsc{AEGEAN}, but we see an excess of faint sources still, due to the different extraction depths. This suggests that applying a beam correction is necessary in order to get accurate flux densities for the sources measured by \textsc{ProFound}. This is something that is intrinsically taken into account when Gaussian components are fit in \textsc{PyBDSF} and \textsc{AEGEAN}. In general though, these beam corrections will only affect the smaller, fainter sources. 

\section{Simulations}
\label{sec:sims}

Although in Section \ref{sec:blindcat} we have shown that \textsc{ProFound} appears to successfully extract accurate flux densities and source morphologies of radio sources, we do not know the true source population in this image. It is therefore hard to quantify whether \textsc{ProFound} can accurately measure the flux densities for all source types in its catalogue. We therefore test on simulated data where the input flux density is known. To do this four variants of simulations are performed. These all make use of the residual image from \textsc{PyBDSF} (where all sources should have been removed and only noise remains). Objects of known flux densities but differing morphologies (in the four different simulations) are then injected into the residual image and recovered. Each simulation performed uses different source morphologies. These are:
\begin{itemize}
\item Gaussian sources with varying sizes
\item Elliptical sources with component sizes from \cite{Wilman2008} convolved with the beam
\item Models of extended sources that \textsc{ProFound} extracted from the original image, these are then re-injected at differing noise levels
\item Extended sources generated from elliptical components from \cite{Wilman2008} convolved with the beam
\end{itemize}

The details of these simulations and the results from each of them are described below. For each simulation, we compare the input and output sources in the same way. To do this we first remove any sources that would be found in the residual image by the different software. This is to ensure that we are not confusing injected sources with sources that could already be detected in the residual image. This is done by performing a positional cross match of the output catalogue from the simulation to the catalogue from running the source extraction software over the residual image with no simulated sources. Sources that are matched within 1" ($\sim 1/5^{\textrm{th}}$ of the PSF) are then removed. Next, we matched the objects in the remaining catalogue to the input sources that were injected into the image, matching within a 3" radius. Finally, we want to consider the possibility that sources in \textsc{PyBDSF} and \textsc{AEGEAN} could consist of multiple components that have not been combined into one source. {For each \textsc{PyBDSF} or \textsc{AEGEAN} source that was not matched to within 3" of an input source, these were investigated to take into account that these could be extra components of a source. This was done through matching these unmatched sources to an input simulated source, provided it was within 20" of the input position of the simulated source.} We also correct the flux density of each source in the \textsc{ProFound} catalogue as in Section \ref{sec:beam_correction}. As the simulations where extended sources from the original image are used will include more extended emission, we use \textsc{PyBDSF} with \texttt{atrous\_do} on to help capture this emission. For the other simulations, we do not use \texttt{atrous\_do} as the emission is smooth. Therefore the extra \texttt{atrous\_do} setting should not be necessary.

\begin{figure*}	
\begin{center}
\centering
\includegraphics[width=14cm]{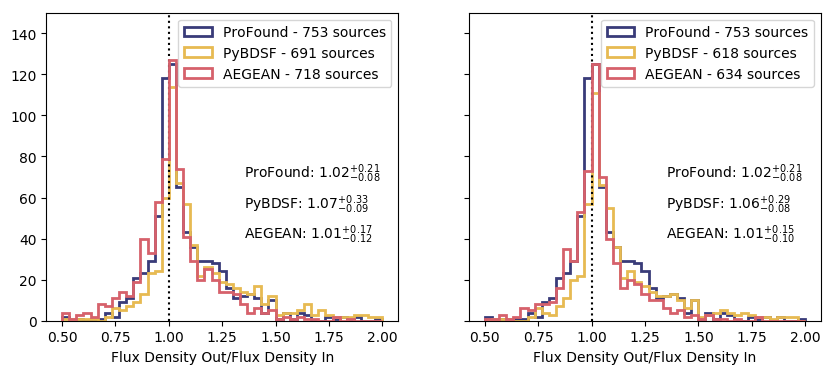}
\caption{{Comparisons of the injected flux densities compared to the fluxes recovered using \textsc{ProFound} (blue), \textsc{PyBDSF} (yellow) and \textsc{AEGEAN} (red). This is for the simulations in which Gaussian sources are injected into the image. Shown is a histogram of the recovered to input flux densities. The median value of the output to input flux density ratio and its uncertainties derived from the 16$^{\textrm{th}}$ and 84$^{\textrm{th}}$ percentiles are shown in both figures and the number of sources detected from each software is shown in the legend. The left hand plots use the source extraction parameters described in Section \ref{sec:detect_param_pybdsf} whilst the right hand plots use the default parameters for \textsc{PyBDSF} and \textsc{AEGEAN}.}}
\label{fig:sims_gaus1}
\end{center}
\end{figure*}

\begin{figure*}	
\begin{center}
\centering
\includegraphics[width=15cm]{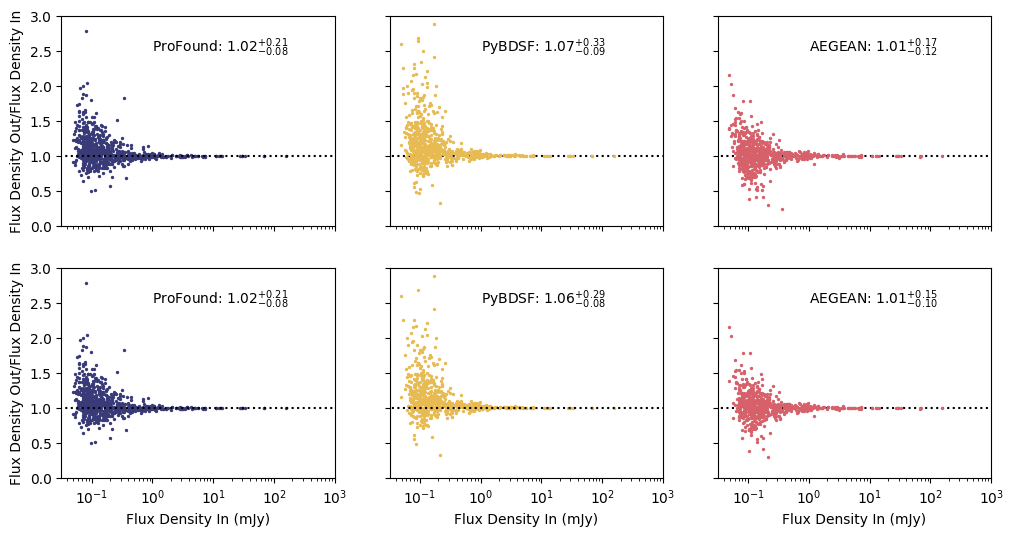}
\caption{{Comparisons of the injected flux densities compared to the fluxes recovered using \textsc{ProFound} (blue, left panel), \textsc{PyBDSF} (yellow, middle panel) and \textsc{AEGEAN} (red, right panel). This is for the simulations in which Gaussian sources are injected into the image. Shown is the ratio of recovered to input fluxes as a function of input flux density. The median value of the output to input flux density ratio and its uncertainties derived from the 16$^{\textrm{th}}$ and 84$^{\textrm{th}}$ percentiles are shown in both figures and the number of sources detected from each software is shown in the legend. The upper row of plots use the source extraction parameters described in Section \ref{sec:detect_param_pybdsf} whilst the lower row of plots use the default parameters for \textsc{PyBDSF} and \textsc{AEGEAN}.}}
\label{fig:sims_gaus2}
\end{center}
\end{figure*}

\begin{figure*}	
\begin{center}
\centering
\includegraphics[width=14cm]{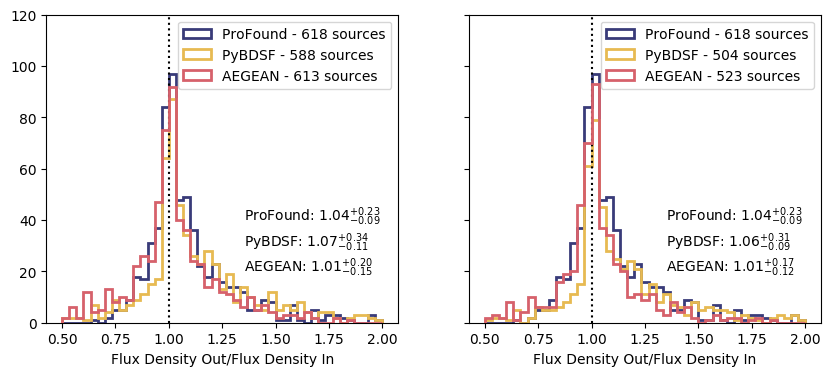}
\caption{{Comparisons of the injected flux densities compared to the fluxes recovered using \textsc{ProFound} (blue), \textsc{PyBDSF} (yellow) and \textsc{AEGEAN} (red). This is for the simulations in which elliptical sources are injected into the image. Shown is a histogram of the recovered to input flux densities. The median value of the output to input flux density ratio and its uncertainties derived from the 16$^{\textrm{th}}$ and 84$^{\textrm{th}}$ percentiles are shown in both figures and the number of sources detected from each software is shown in the legend. The left hand plots use the source extraction parameters described in Section \ref{sec:detect_param_pybdsf} whilst the right hand plots use the default parameters for \textsc{PyBDSF} and \textsc{AEGEAN}.}}
\label{fig:sims_ellipse1}
\end{center}
\end{figure*}

\begin{figure*}	
\begin{center}
\centering
\includegraphics[width=15cm]{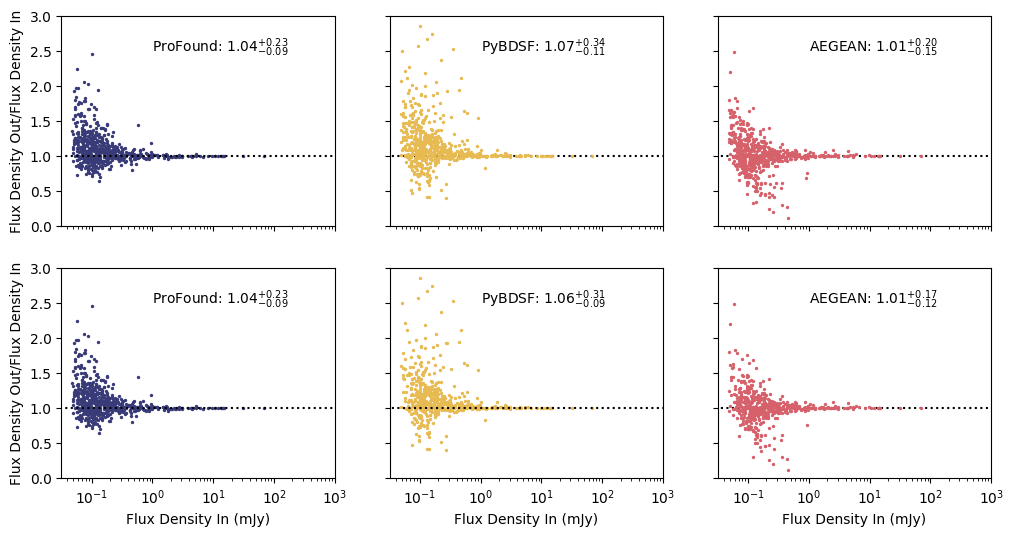}
\caption{{Comparisons of the injected flux densities compared to the fluxes recovered using \textsc{ProFound} (blue, left panel), \textsc{PyBDSF} (yellow, middle panel) and \textsc{AEGEAN} (red, right panel). This is for the simulations in which elliptical sources are injected into the image. Shown is the ratio of recovered to input fluxes as a function of input flux density. The median value of the output to input flux density ratio and its uncertainties derived from the 16$^{\textrm{th}}$ and 84$^{\textrm{th}}$ percentiles are shown in both figures and the number of sources detected from each software is shown in the legend. The upper row of plots use the source extraction parameters described in Section \ref{sec:detect_param_pybdsf} whilst the lower row of plots use the default parameters for \textsc{PyBDSF} and \textsc{AEGEAN}.}}
\label{fig:sims_ellipse2}
\end{center}
\end{figure*}

\begin{figure*}	
\begin{center}
\centering
\includegraphics[width=14cm]{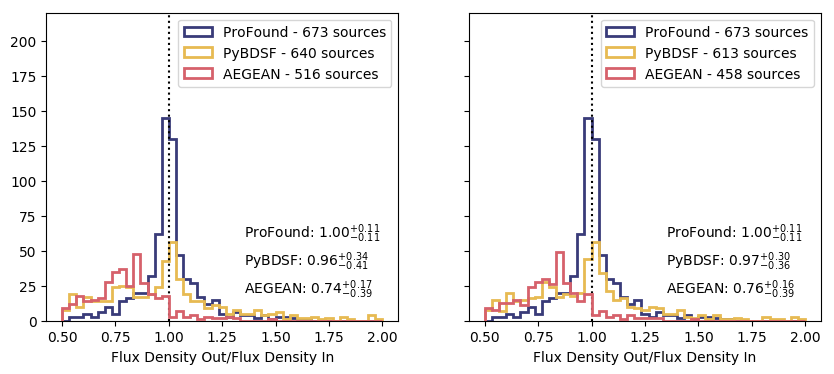}
\caption{{Comparisons of the injected flux densities compared to the fluxes recovered using \textsc{ProFound} (blue), \textsc{PyBDSF} (with \texttt{atrous\_do}, yellow) and \textsc{AEGEAN} (red). This is for the simulations in which extended objects extracted from the original image using \textsc{ProFound} are injected. Shown is a histogram of the recovered to input flux densities. The median value of the output to input flux density ratio and its uncertainties derived from the 16$^{\textrm{th}}$ and 84$^{\textrm{th}}$ percentiles are shown in both figures and the number of sources detected from each software is shown in the legend. The left hand plots use the source extraction parameters described in Section \ref{sec:detect_param_pybdsf} whilst the right hand plots use the default parameters for \textsc{PyBDSF} and \textsc{AEGEAN}.}}
\label{fig:sims_extended_atrous1}
\end{center}
\end{figure*}

\begin{figure*}	
\begin{center}
\centering
\includegraphics[width=15cm]{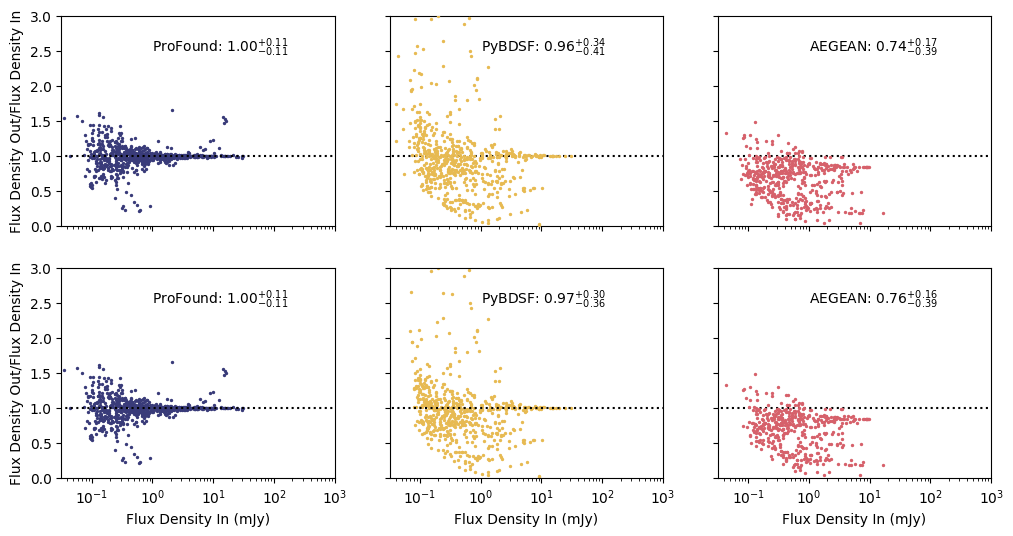}
\caption{{Comparisons of the injected flux densities compared to the fluxes recovered using \textsc{ProFound} (blue, left panel), \textsc{PyBDSF} (yellow, middle panel) and \textsc{AEGEAN} (red, right panel). This is for the simulations in which extended objects extracted from the original image using \textsc{ProFound} are injected. Shown is the ratio of recovered to input fluxes as a function of input flux density. The median value of the output to input flux density ratio and its uncertainties derived from the 16$^{\textrm{th}}$ and 84$^{\textrm{th}}$ percentiles are shown in both figures and the number of sources detected from each software is shown in the legend. The upper row of plots use the source extraction parameters described in Section \ref{sec:detect_param_pybdsf} whilst the lower row of plots use the default parameters for \textsc{PyBDSF} and \textsc{AEGEAN}.}}
\label{fig:sims_extended_atrous2}
\end{center}
\end{figure*}

\begin{figure*}	
\begin{center}
\centering
\includegraphics[width=14cm]{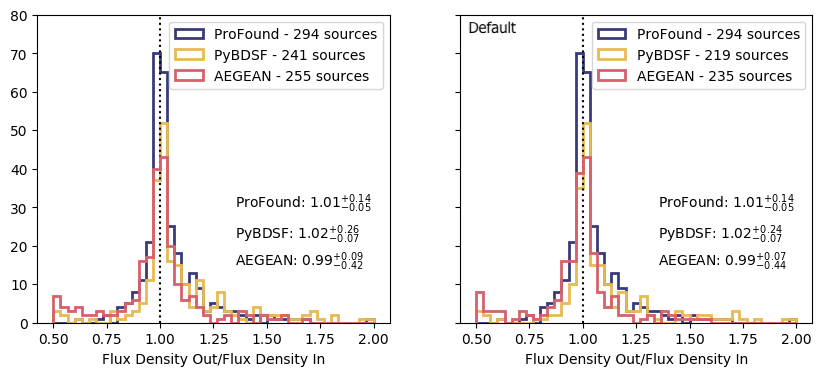}
\caption{{Comparisons of the injected flux densities compared to the fluxes recovered using \textsc{ProFound} (blue), \textsc{PyBDSF} (without \texttt{atrous\_do}, yellow) and \textsc{AEGEAN} (red). This is for the simulations in which multi-component elliptical sources are injected into the image. Shown is a histogram of the recovered to input flux densities. The median value of the output to input flux density ratio and its uncertainties derived from the 16$^{\textrm{th}}$ and 84$^{\textrm{th}}$ percentiles are shown in both figures and the number of sources detected from each software is shown in the legend. The left hand plots use the source extraction parameters described in Section \ref{sec:detect_param_pybdsf} whilst the right hand plots use the default parameters for \textsc{PyBDSF} and \textsc{AEGEAN}.}}
\label{fig:sims_skads1}
\end{center}
\end{figure*}

\begin{figure*}	
\begin{center}
\centering
\includegraphics[width=15cm]{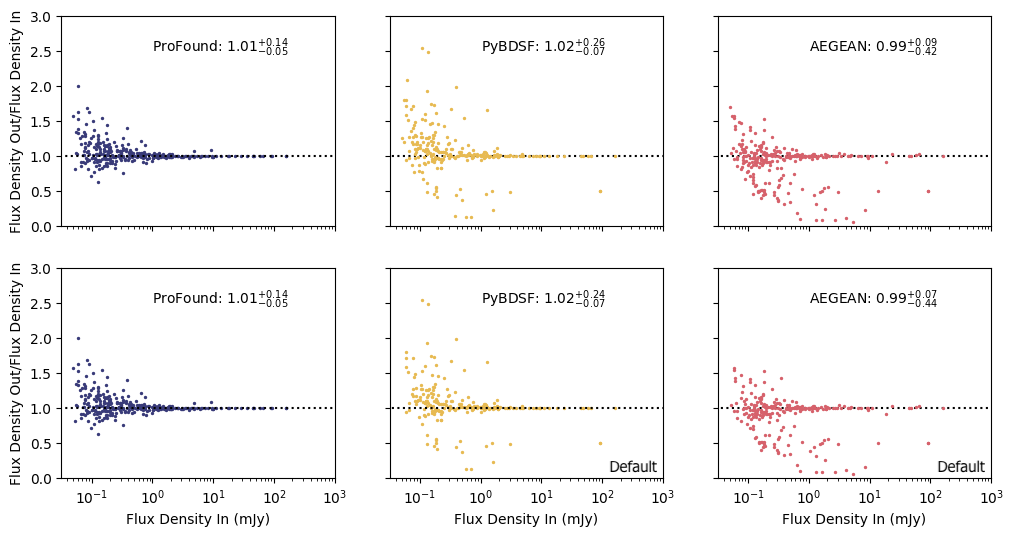}
\caption{{Comparisons of the injected flux densities compared to the fluxes recovered using \textsc{ProFound} (blue, left panel), \textsc{PyBDSF} (yellow, middle panel) and \textsc{AEGEAN} (red, right panel). This is for the simulations in which multi-component elliptical sources are injected into the image. Shown is the ratio of recovered to input fluxes as a function of input flux density. The median value of the output to input flux density ratio and its uncertainties derived from the 16$^{\textrm{th}}$ and 84$^{\textrm{th}}$ percentiles are shown in both figures and the number of sources detected from each software is shown in the legend. The upper row of plots use the source extraction parameters described in Section \ref{sec:detect_param_pybdsf} whilst the lower row of plots use the default parameters for \textsc{PyBDSF} and \textsc{AEGEAN}.}}
\label{fig:sims_skads2}
\end{center}
\end{figure*}

\begin{figure*}	
\begin{center}
\begin{minipage}[b]{\textwidth}
\centering
\includegraphics[width=0.7\textwidth]{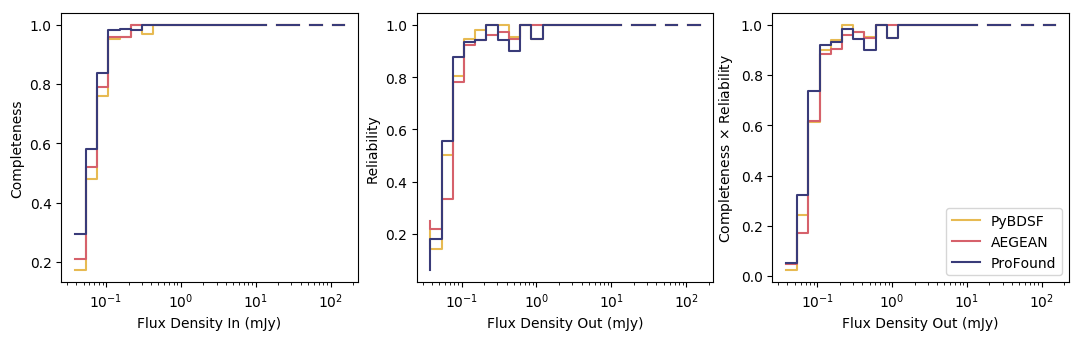}
\subcaption{Completeness (left), Reliability (centre) and Completeness$\times$Reliability (right) for Gaussian simulations}
\end{minipage}%
\newline
\begin{minipage}[b]{\textwidth}
\centering
\includegraphics[width=0.7\textwidth]{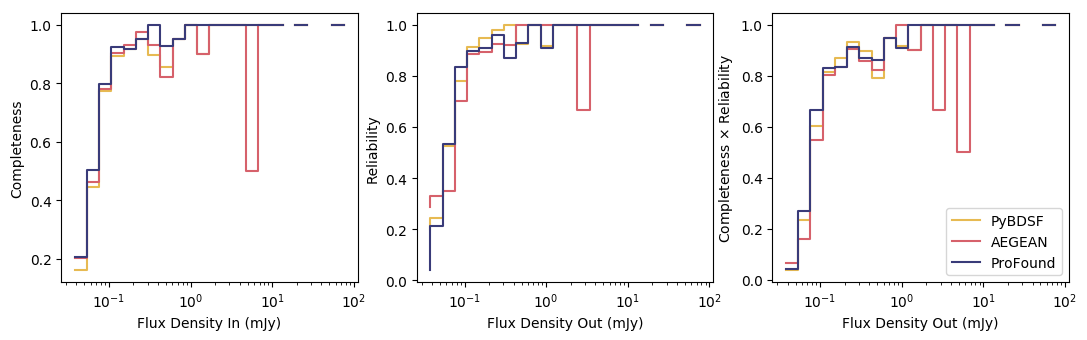}
\subcaption{Completeness (left), Reliability (centre) and Completeness$\times$Reliability (right) for elliptical sources simulations}
\end{minipage}%
\newline
\begin{minipage}[b]{\textwidth}
\centering
\includegraphics[width=0.7\textwidth]{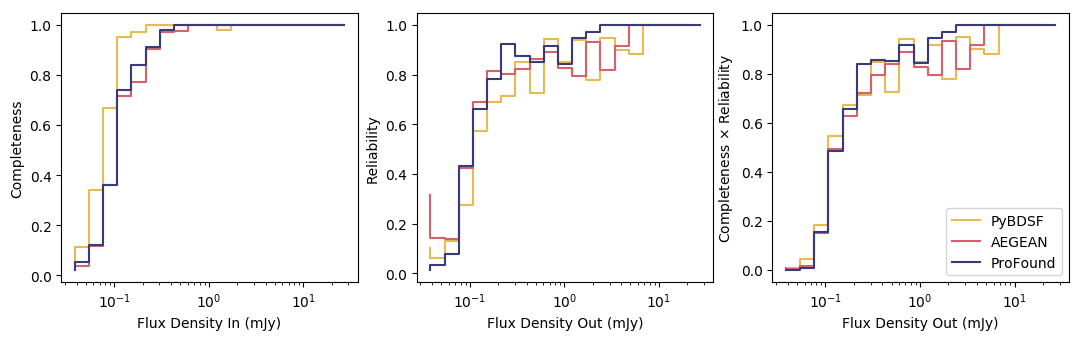}
\subcaption{Completeness (left), Reliability (centre) and Completeness$\times$Reliability (right) for simulations with extended objects extracted from the original image with \textsc{ProFound}}
\end{minipage}%
\newline
\begin{minipage}[b]{\textwidth}
\centering
\includegraphics[width=0.7\textwidth]{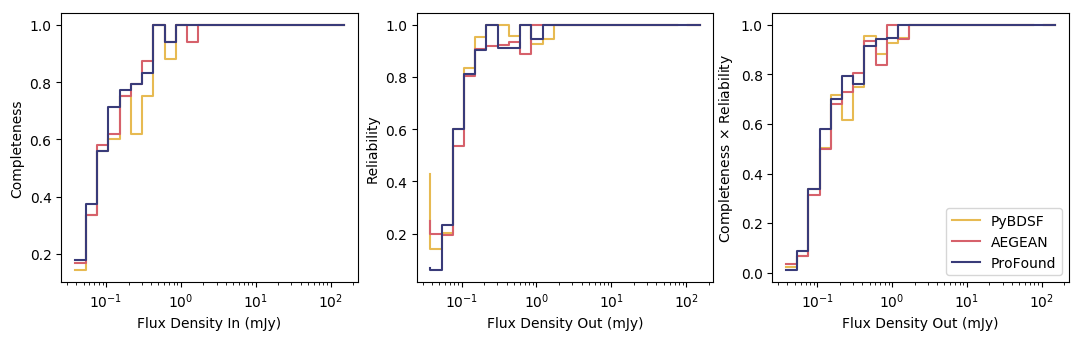}
\subcaption{Completeness (left), Reliability (centre) and Completeness$\times$Reliability (right) for multi-component elliptical sources simulations}
\end{minipage}%
\caption{{Comparisons of the completeness (left hand plots), reliability (central plots) and completeness$\times$reliability (right hand plots) using \textsc{ProFound} (blue), \textsc{PyBDSF} (yellow) and \textsc{AEGEAN} (red) for the four simulations performed.  The completeness and reliability for the simulations with: (a) Gaussian sources; (b) elliptical sources; (c) extended objects extracted from the original image with \textsc{ProFound} and (d) multi-component elliptical sources are shown. These are shown using the detection parameters of \textsc{PyBDSF} and \textsc{AEGEAN} determined in Section \ref{sec:detect_param_pybdsf} but with \texttt{atrous\_do} used for \textsc{PyBDSF} in (c).}}
\label{fig:completeness_reliability}
\end{center}
\end{figure*}

\subsection{Gaussian Sources}
\label{sec:sims_gauss}
Firstly simulated Gaussian sources are injected into the image. In order to not be limited to unresolved objects, a range of sizes are generated using the observed sizes from the \textsc{PyBDSF} catalogue. The major and minor axes sizes from this catalogue are modelled as a normalised histogram from which a process of sampling is used to generate {major and minor axes} of the simulated Gaussians. A random orientation is also assigned to these Gaussian components. A flux density is assigned to each source using the SKA Simulated Skies \citep[S$^3$; ][]{Wilman2008}. These are semi-empirical simulations of the radio sky and provide realistic distributions of expected source counts at 5 radio frequencies. We make use of the 1.4 GHz flux densities and randomly assign each Gaussian component a flux density from this. We only include those sources with total S$^3$ flux densities equivalent to point sources with a peak flux density $>3\sigma_{map}$, where $\sigma_{map}$ is the typical noise in the image which is taken as 16 $\mu$Jy/beam. 

Combining the random flux density from S$^3$ with the major/minor axes sampled from the source distribution, we then generate 1000 Gaussians that are added into the residual image. When adding in the simulated Gaussians, a record of their positions, sizes and flux densities (both from S$^3$ and from summing the injected flux density per beam in the injected pixels) is recorded. \textsc{ProFound}, \textsc{PyBDSF} and \textsc{AEGEAN} are then run over the simulated image. {We compare the flux densities of each Gaussian source in the \textsc{PyBDSF} and \textsc{AEGEAN} catalogue to the S$^3$ flux used for the input source. For the \textsc{ProFound} catalogue, the measured flux densities are compared to the flux density from the sum of the injected pixels for a like-to-like comparison}\footnote{The difference in these flux densities is typically negligible (<1\%) and arise from whether the Gaussian is centred in a pixel when injected into the image. We use this like-to-like flux density comparison for the simulations in Sections \ref{sec:sims_ellipse} and \ref{sec:sims_skads} as well}. 

{The results of comparing the ratio of the recovered flux density to the injected flux density can be seen in Figures \ref{fig:sims_gaus1} and \ref{fig:sims_gaus2}. In these we also record the median ratio as well as {uncertainties} generated from the 16$^{\textrm{th}}$ and 84$^{\textrm{th}}$ percentiles. With \textsc{ProFound} and \textsc{AEGEAN} we find ratios of $\sim$1, with \textsc{PyBDSF} having a slightly larger median value. \textsc{ProFound} gives a ratio of {$1.02^{+0.21}_{-0.08}$} compared to \textsc{PyBDSF} which gives a ratio of $\sim 1.06$ and \textsc{AEGEAN} with a ratio of $\sim 1.01$. The scatter with \textsc{ProFound}, \textsc{PyBDSF} and \textsc{AEGEAN} are all comparable, with all showing an excess towards higher flux ratios.  }

{All three show that they are doing a successful job at modelling the total emission from Gaussian sources. \textsc{ProFound} finds the most sources compared to \textsc{PyBDSF} and \textsc{AEGEAN}, however only by $\sim$ 25 sources compared to \textsc{AEGEAN}. The extra sources are due to the differences in $\sigma$ levels necessary to be classified as a source. Despite the difference in $\sigma$ levels used for extracting sources, \textsc{ProFound} is still capable at these lower noise levels of, on average, accurately recovering the flux densities, as can be seen in Figure \ref{fig:sims_gaus2}. Overall, our results suggests \textsc{ProFound} is comparable with \textsc{PyBDSF} and \textsc{AEGEAN} of being used as a source extractor for Gaussian-like objects. }

\subsection{Elliptical sources}
\label{sec:sims_ellipse}
For our next simulation, we consider the scenario where radio sources are not intrinsic Gaussians and instead are disc-like objects of uniform brightness that are convolved with the beam. This will deal with the question of how well both \textsc{ProFound} and the other source extractors model sources that are not inherently Gaussian. 

To do this, we use the S$^3$ components catalogue which includes information on the {major and minor axes} of the elliptical components used. A component is randomly selected and the sizes and corresponding flux density for this component is then used to model the source as an ellipse of uniform surface brightness. We use the same flux density limit as in Section \ref{sec:sims_gauss}. This is then convolved with the Gaussian restoring beam of the radio observations. Again 1000 simulated sources are injected into the residual image and the extracted catalogue is compared to the injected sources in the same way as in Section \ref{sec:sims_gauss}.

{The results from this simulation are shown in Figures \ref{fig:sims_ellipse1} and \ref{fig:sims_ellipse2}. Again all the source extraction software exhibit peaks around $\sim1$ for the ratio of the recovered to the injected flux density. However there is typically an excess at high ratios. Whereas \textsc{AEGEAN} has a median ratio of $\sim 1.01$, \textsc{PyBDSF} and \textsc{ProFound} both appear to find an excess of emission compared to what is in injected, $\sim 1.04$ for \textsc{ProFound} and 1.06 for \textsc{PyBDSF} and therefore may be slightly over predicting the flux density of a source. All three however give peaks around the same value and have similar scatter to one another, demonstrating that they all perform similarly well for the simple uniform elliptical source morphology.}

\subsection{Extended Sources - from the image}
\label{sec:sims_extended}
Next, we focus on sources that consist of complicated morphologies. To do this, we use the large objects found in the \textsc{ProFound} blind detection of the VLA image and re-inject these in the residual image. These large sources were typically the most complicated morphologies. We define `large' here as those that had {an \texttt{R50}$\ge$3.5, where \texttt{R50} is defined} in \textsc{ProFound} as the approximate elliptical semi-major axis containing 50\% of the flux. {This corresponds to 81 objects within the central $\sim$1deg$2$.}

{To avoid any issues of this becoming a circular argument where we extract radio emission from \textsc{ProFound} and then re-extract using \textsc{ProFound} to see how well \textsc{ProFound} behaves, we artificially multiply the models of the sources by a random factor. This factor is generated as a random number between 0.01-1.0 but selected so that it is sampled uniformly in logarithmic space. By doing this and by {injecting these sources at random positions,} we change the effect of the noise. This is likely to make it more difficult to extract with \textsc{ProFound}. For these simulations we add in each object 5 times to give a total of 405 sources in our input catalogue. Fewer sources were used due to the extended nature of these sources and so in order to avoid sources being merged together, their numbers were reduced}.

{Due to the lower recovery rate of sources and the complicated nature of the sources themselves, the simulations were repeated 5 times and the combined results of these are shown in Figures \ref{fig:sims_extended_atrous1} and \ref{fig:sims_extended_atrous2}. From Figures \ref{fig:sims_extended_atrous1} and \ref{fig:sims_extended_atrous2}, it is evident that \textsc{ProFound} does an excellent job recovering the flux densities of sources compared to \textsc{PyBDSF} and \textsc{AEGEAN}. In these simulations, \textsc{ProFound} gives a flux density ratio of $1.00^{+0.11}_{-0.11}$ whereas \textsc{AEGEAN} has a ratio of $0.76^{+0.16}_{-0.39}$, when the default parameters are used. This shows that \textsc{AEGEAN} is underestimating the flux density of objects that have complicated and large morphologies. As using the \texttt{atrous\_do} mode will be important in this simulation, the results from \textsc{PyBDSF} using this is shown in Figures \ref{fig:sims_extended_atrous1} and \ref{fig:sims_extended_atrous2}. The results from \textsc{PyBDSF} are centred on a value of 1 ($0.97^{+0.30}_{-0.36}$), suggesting \textsc{PyBDSF} is able to accurately recover the emission from extended sources. However the scatter is much larger than for \textsc{ProFound}, with values of $\sim 0.3-4$ for the scatter with \textsc{PyBDSF} compared to $\sim0.1$ for \textsc{ProFound}. This suggests both \textsc{PyBDSF} and \textsc{AEGEAN} may struggle to consistently model the entire emission of the source or that it may be harder to combine multiple components together in a consistent way (as it is done here purely within a fixed angular separation here). This suggests that in previous continuum surveys the flux densities of complicated sources may have been under/over estimated. This has implications for the descriptions of radio source populations, such as source counts, luminosity functions and spectral indices. \textsc{ProFound} also has a much smaller scatter in the flux density ratios that it calculates compared to \textsc{PyBDSF} and \textsc{ProFound}. This emphasises \textsc{ProFound}'s ability to accurately extract the flux densities of those source with complex morphologies.}

\subsection{Extended sources - multi-component elliptical sources}
\label{sec:sims_skads}
For our final simulations, we again investigate how well extended sources can be recovered, this time using the component catalogues of S$^3$ \citep{Wilman2008}. In Section \ref{sec:sims_ellipse}, we injected elliptical components from S$^3$ convolved with the beam, however these were single individual components. In this simulation, we instead inject all components of one source into the residual image. In S$^3$, Star Forming Galaxies (SFGs) are described as one component objects, where as FRI and FRII type Active Galactic Nuclei (AGN) are formed of multiple components of e.g. a core, jets and hotspots. These are all described by elliptical components, which we convolve with the beam individually, before summing together to form the source. We inject 500 of these multi-component objects. {Again, fewer sources are injected due to the extended nature of these sources and we also do not inject single component sources, i.e. SFGs, as these are the same as from the simulations in Section \ref{sec:sims_ellipse}. }

{The results of our recovered to injected flux density ratios can be seen in Figures \ref{fig:sims_skads1} and  \ref{fig:sims_skads2}.} This suggests that all three detection mechanisms seem to do a good job in re-extracting the flux density of these sources, with all having a flux density ratio of $\sim$1. This is a value of $1.01^{+0.14}_{-0.05}$ for \textsc{ProFound}, $1.02^{+0.24}_{-0.07}$ for \textsc{PyBDSF} and $0.99^{+0.07}_{-0.44}$ for \textsc{AEGEAN}, again using the default parameters. This suggests that all three of these source extractors are able to sensibly model objects that have smooth, double-lobed morphologies. However again the scatter in \textsc{ProFound} is typically much smaller than for \textsc{PyBDSF} or \textsc{AEGEAN}, suggesting \textsc{ProFound} can more often recover the flux densities of these sources accurately.  \\

\noindent Overall these simulations suggest that \textsc{PyBDSF} and \textsc{AEGEAN} perform well for most source types however are less suitable to extract the emission of radio sources that have complex morphologies. \textsc{ProFound} however has shown that it is capable of successfully determining the flux densities for a variety of source morphologies, including the Gaussians and sources with complicated morphologies that are typically observed in radio continuum observations. 

\subsection{Completeness and Reliability}
{We also show, in Figure \ref{fig:completeness_reliability}, the completeness and reliability distribution as a function of flux density for each of the simulations discussed in Sections \ref{sec:sims_gauss} to \ref{sec:sims_skads}. Completeness is defined as the fraction of sources that are input into the simulated images for which the source is found in the output catalogue. Reliability on the other hand is the fraction of sources obtained in the output catalogue of the simulation that have a counterpart in the input catalogue.}

{To determine completeness and reliability, the input and output catalogues were matched within an angular radius. For both of these, only sources that had RA/Dec values within the central deg$^2$ of the image (i.e. the overlap region with CFHTLS D1) were considered, this was to ensure that any noise detection from around the region of higher rms around outside of the image were not included, as sources were only detected in this central region. The angular radius used here is given as 3" (as used earlier in Sections \ref{sec:sims_gauss} to \ref{sec:sims_skads}) for the Gaussian and Elliptical simulations (Figures \ref{fig:completeness_reliability}(a) and (b), as these are compact, smooth sources. For the extended objects, described in Sections \ref{sec:sims_extended} and \ref{sec:sims_skads} (Figures \ref{fig:completeness_reliability}(c) and (d)), due to the larger nature of these objects, and the multi-component nature of the objects described in Section \ref{sec:sims_skads}, a larger angular radius is used. This is taken to be 15" (or $\sim 3 \times$ the beam size). As well as showing Completeness (left hand panels) and Reliability (central panels), we also present the product of the two: Completeness$\times$Reliability (right hand panels), this is to indicate a compromise between the two. }

{Figure \ref{fig:completeness_reliability} shows that \textsc{ProFound} has comparable Completeness, Reliability and Completeness$\times$Reliability to both \textsc{PyBDSF} and \textsc{AEGEAN}, demonstrating that it is comparable to other known radio source extractors, despite its different approach to extracting sources. For Figure \ref{fig:completeness_reliability}, the parameters determined in Section \ref{sec:detect_param_pybdsf} are used for \textsc{AEGEAN} and \textsc{PyBDSF} to minimize the effect of different false detection levels on reliability. For extended sources (Figure \ref{fig:completeness_reliability}(c)), however \textsc{ProFound} produces slightly larger values of completeness$\times$reliability compared to \textsc{PyBDSF} and \textsc{AEGEAN} for extended sources over $\sim 0.2-5$mJy. This suggests that \textsc{ProFound} is successfully modelling this complicated emission. However, these will all be influenced by the matching radius used as well as whether sources have been merged together into a single source or not, or whether it has been split into multiple components, both of which can put the positions of the new sources at large distances from the original location of the source(s).  Therefore, this should be taken into account when considering the plots shown in Figure \ref{fig:completeness_reliability}.}

\begin{figure*}
\begin{center}
\begin{minipage}[b]{\textwidth}
\centering
\includegraphics[width=15cm]{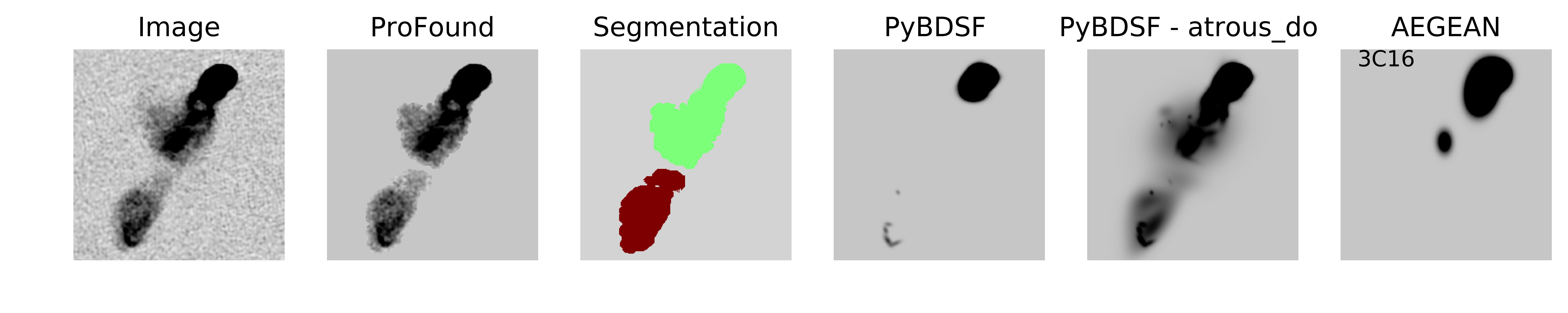}
\end{minipage}%
\newline
\begin{minipage}[b]{\textwidth}
\centering
\includegraphics[width=15cm]{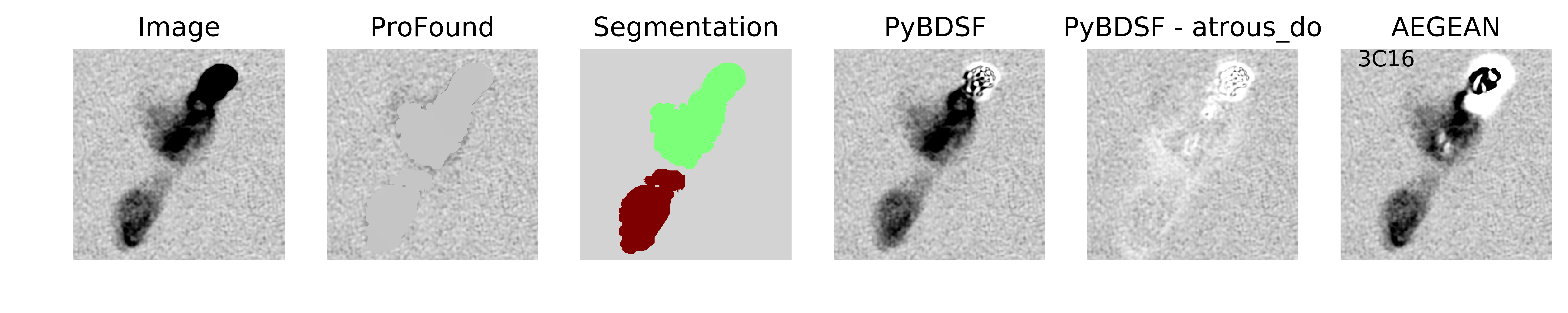}
\subcaption{3C16}
\end{minipage}%
\newline \newline
\begin{minipage}[b]{\textwidth}
\centering
\includegraphics[width=15cm]{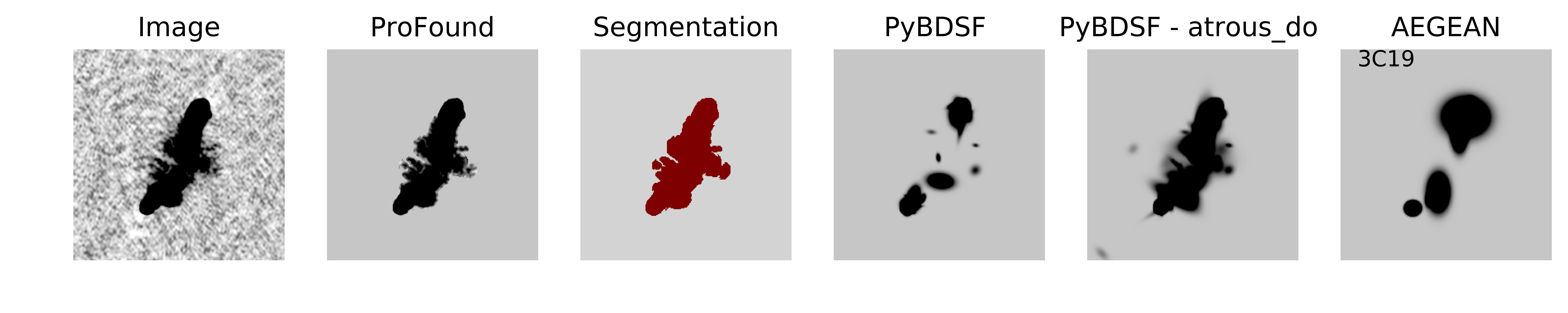}
\end{minipage}%
\newline 
\begin{minipage}[b]{\textwidth}
\centering
\includegraphics[width=15cm]{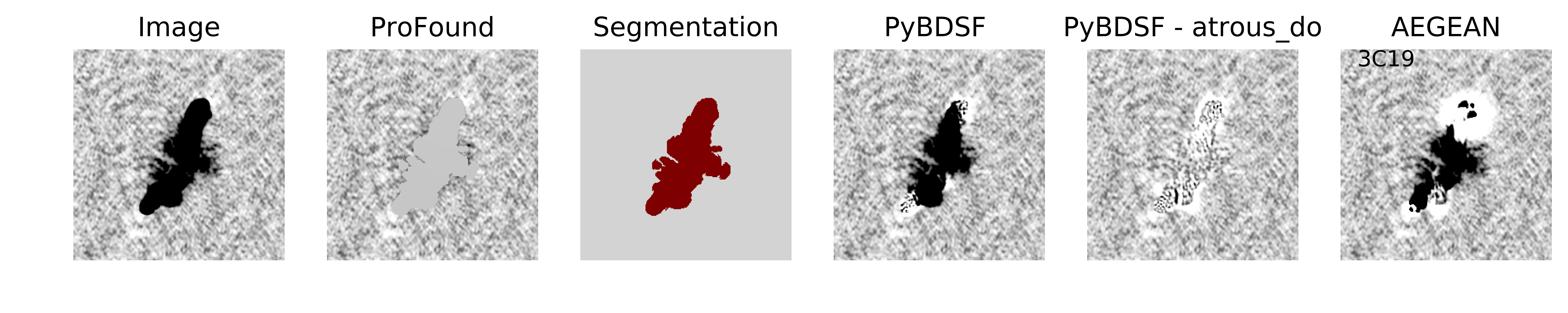}
\subcaption{3C19}
\end{minipage}%
\newline \newline
\begin{minipage}[b]{\textwidth}
\centering
\includegraphics[width=15cm]{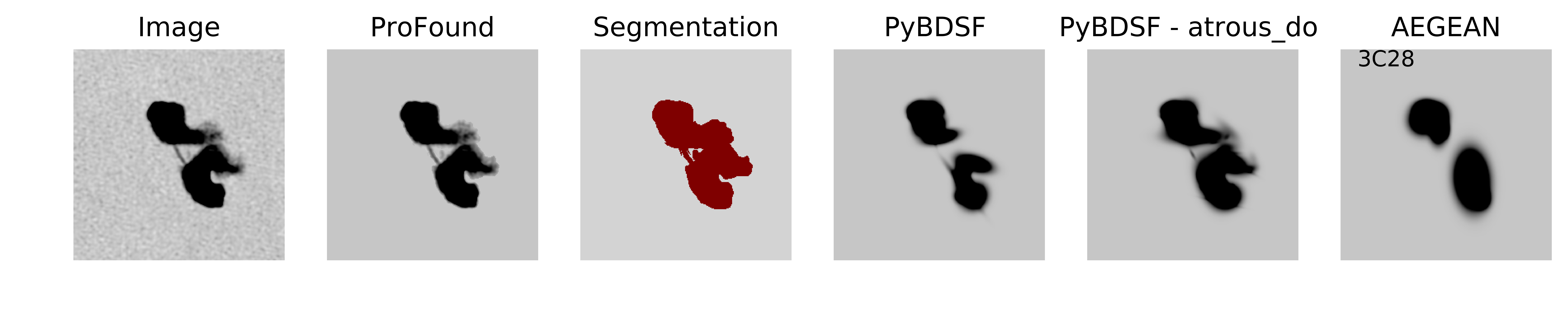}
\end{minipage}%
\newline 
\begin{minipage}[b]{\textwidth}
\centering
\includegraphics[width=15cm]{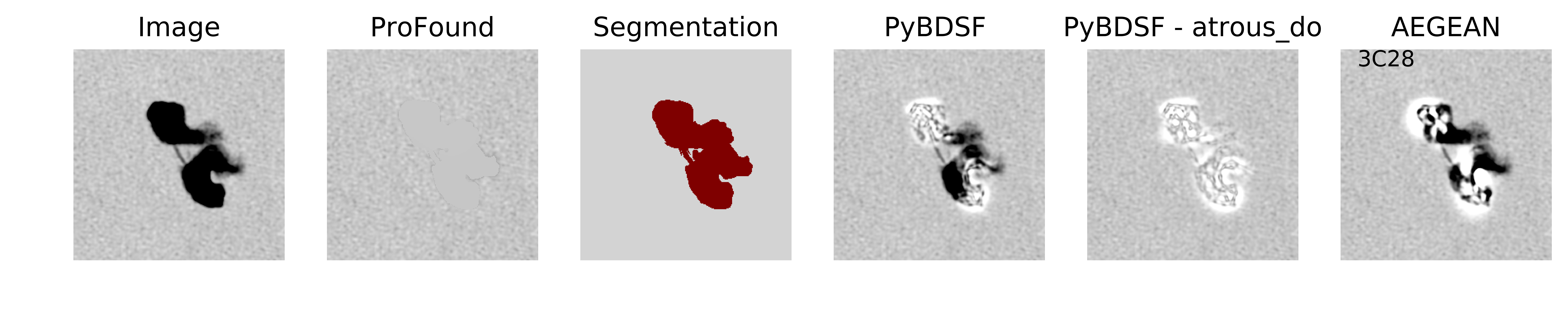}
\subcaption{3C28}
\end{minipage}%
\caption{Comparison of the model (top row) and residual (bottom row) images from \textsc{ProFound} (panel 2), \textsc{PyBDSF} (panel 4 without \texttt{atrous\_do} and panel 5 with \texttt{atrous\_do}=True) and \textsc{AEGEAN} (panel 6) of five 3C sources (whose images are shown in panel 1). These are shown for the sources: 3C16 (a), 3C19 (b), 3C28 (c), 3C42 (d) and 3C47 (e). The segmentation images from \textsc{ProFound} are shown in panel 3. Here \textsc{PyBDSF} and \textsc{AEGEAN} use the default parameters as described in Section \ref{sec:detect_param}, and used in \protect \cite{Hopkins2015}. Shown are the entirety of the images downloaded from \protect \cite{Leahy1996ATLAS}.}
\end{center}
\end{figure*}

\begin{figure*}
\ContinuedFloat
\begin{center}
\begin{minipage}[b]{\textwidth}
\centering
\includegraphics[width=15cm]{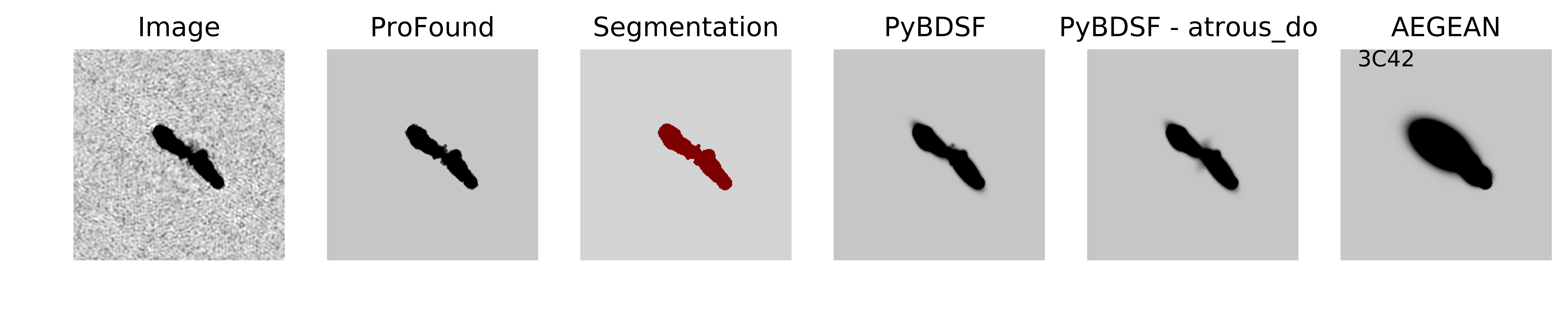}
\end{minipage}%
\newline 
\begin{minipage}[b]{\textwidth}
\centering
\includegraphics[width=15cm]{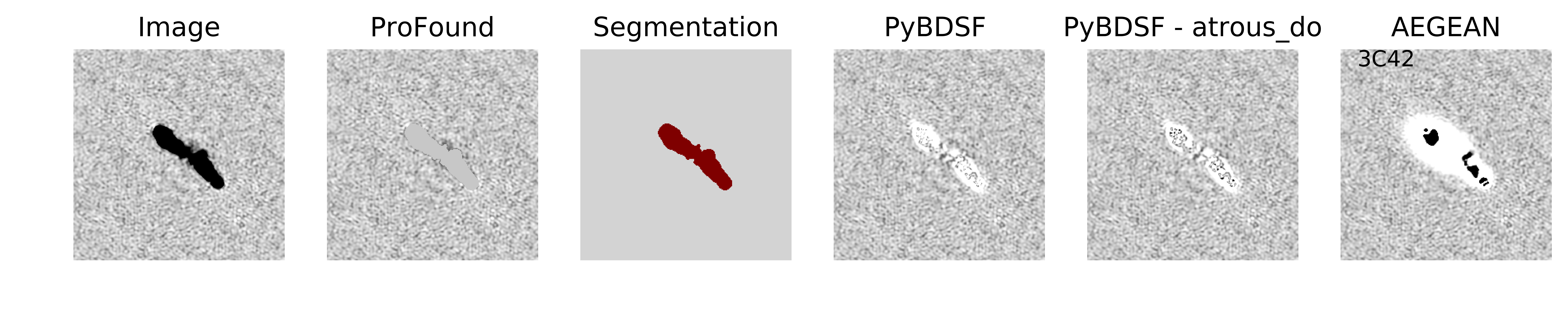}
\subcaption{3C42}
\end{minipage}%
\newline \newline
\begin{minipage}[b]{\textwidth}
\centering
\includegraphics[width=15cm]{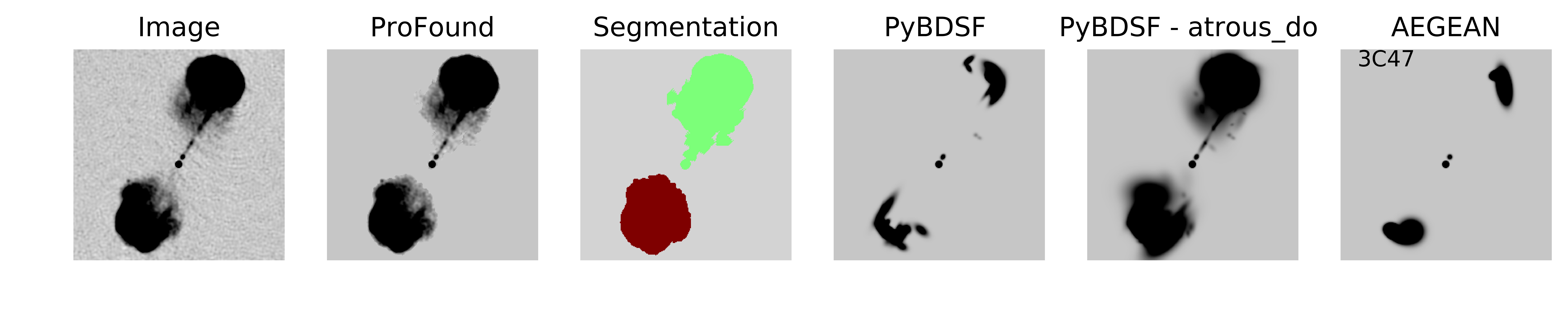}
\end{minipage}%
\newline 
\begin{minipage}[b]{\textwidth}
\centering
\includegraphics[width=15cm]{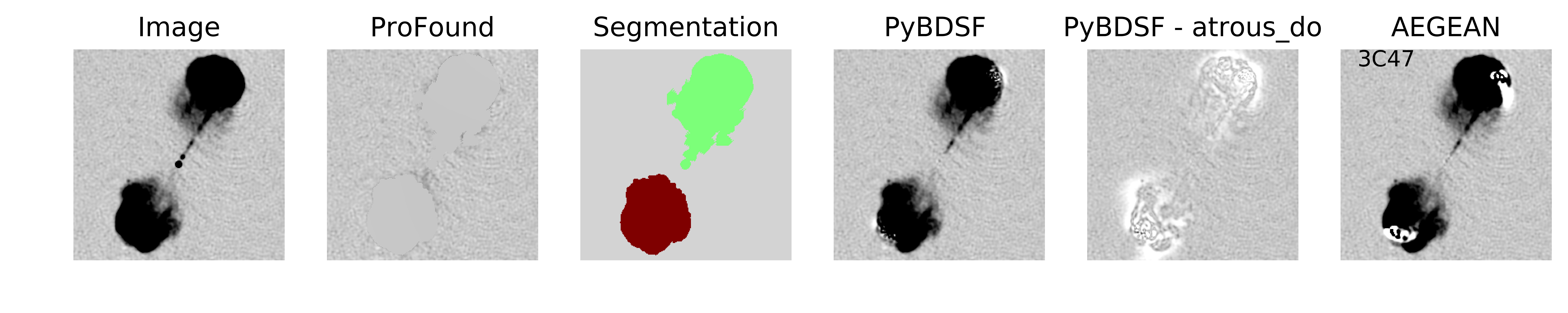}
\subcaption{3C47}
\end{minipage}%
\caption{Continued ... Comparison of the model (top row) and residual (bottom row) images from \textsc{ProFound} (panel 2), \textsc{PyBDSF} (panel 4 without \texttt{atrous\_do} and panel 5 with \texttt{atrous\_do}=True) and \textsc{AEGEAN} (panel 6) of five 3C sources (whose images are shown in panel 1). These are shown for the sources: 3C16 (a), 3C19 (b), 3C28 (c), 3C42 (d) and 3C47 (e). The segmentation images from \textsc{ProFound} are shown in panel 3. {Here \textsc{PyBDSF} and \textsc{AEGEAN} use the default parameters as described in Section \ref{sec:detect_param}, and used in \protect \cite{Hopkins2015}. Shown are the entirety of the images downloaded from \protect \cite{Leahy1996ATLAS}. }}
\label{fig:3csources}
\end{center}
\end{figure*}

\section{3C Sources}
\label{sec:3C}

As the benefits of \textsc{ProFound} arise in its ability to model and calculate flux densities of extended sources, we make one final comparison to compare how well the different software perform on known extended sources. To do this we use images of five 3C sources that were described in Section \ref{sec:data_3C}. To compare the fitting of the sources, we both compared the visual fitting as well as investigating the residuals of the image, as in Section \ref{sec:blindcat}.

A comparison of the visual models of these sources from the different source extractors can be seen in Figure \ref{fig:3csources} (top panel for each source). The image is shown on the left hand side of each sub-figure panel with the models from \textsc{ProFound}, \textsc{PyBDSF} and \textsc{AEGEAN} also shown. We also include the segmentation map from \textsc{ProFound} of the source. Each colour in these plots represents a different source as defined by \textsc{ProFound}, after the grouping mentioned in Section \ref{sec:detect_param_profound}. As with the other comparison images, the images from \textsc{PyBDSF} and \textsc{AEGEAN} here are those using the default extraction parameters. The bottom panel for each object shows the corresponding residual image.

Figure \ref{fig:3csources} illustrates how \textsc{ProFound} is able to trace the shape of the source and so model its radio emission. In the cases shown here, both \textsc{PyBDSF} (without \texttt{atrous\_do}) and \textsc{AEGEAN} do not adequately model the emission seen in the image. Visually, they are unable to constrain the complicated morphology of these sources. For components that are missing, many of these are bright, compared to the sky level, and so it is not a $\sigma$ level discrepancy that causes components to be missing or not well modelled. With \texttt{atrous\_do}, however, \textsc{PyBDSF} is able to better model the emission of these sources. The residual images in Figure \ref{fig:3csources} also show how \textsc{ProFound} is tracing the shape well but also includes some noise in the model of the source. For \textsc{PyBDSF} and \textsc{AEGEAN} the 3C images appear to have been over-fit in areas, which can leave negative residuals around the source. 

\begin{figure*}
\begin{center}
\begin{minipage}[b]{0.5\textwidth}
\centering
\includegraphics[width=8.9cm]{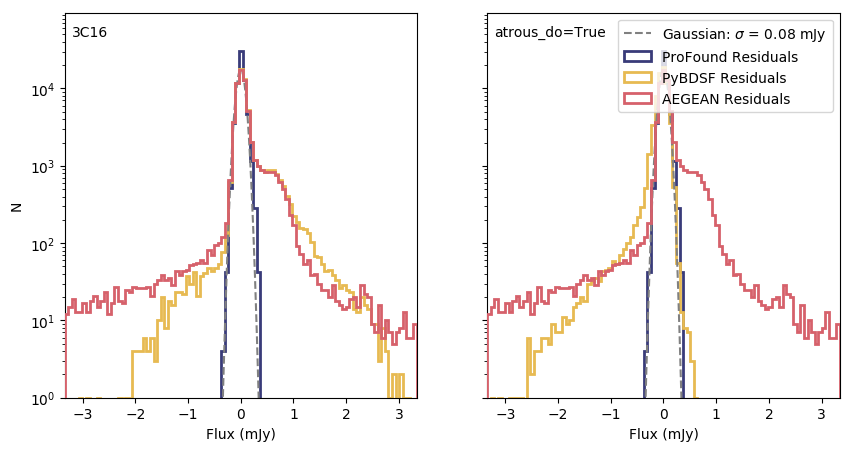}
\subcaption{3C16}
\end{minipage}%
\begin{minipage}[b]{0.5\textwidth}
\centering
\includegraphics[width=8.9cm]{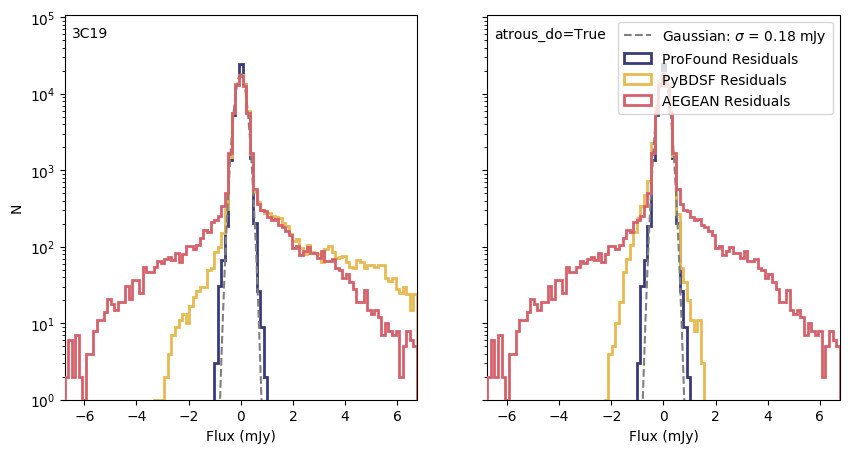}
\subcaption{3C28}
\end{minipage}%
\newline
\begin{minipage}[b]{0.5\textwidth}
\centering
\includegraphics[width=8.9cm]{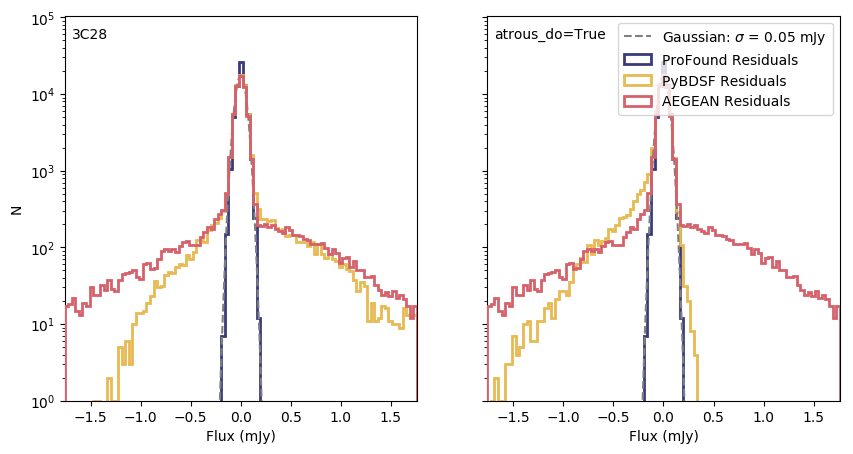}
\subcaption{3C28}
\end{minipage}%
\begin{minipage}[b]{0.5\textwidth}
\centering
\includegraphics[width=8.9cm]{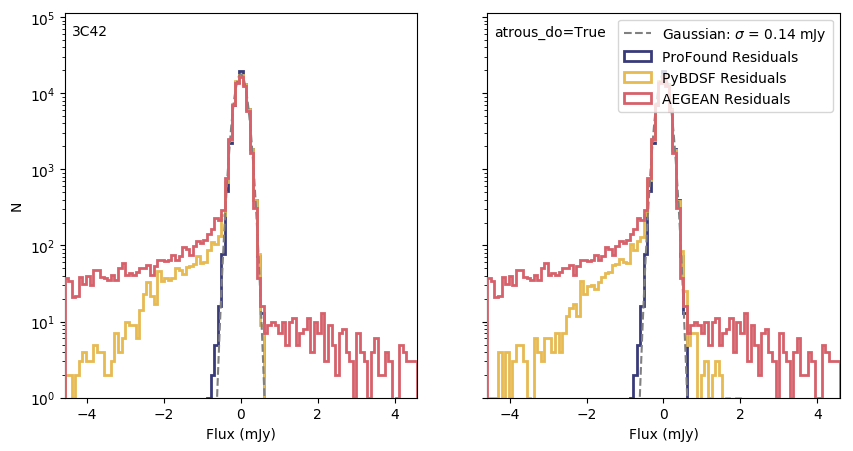}
\subcaption{3C42}
\end{minipage}%
\newline
\begin{minipage}[b]{0.5\textwidth}
\centering
\includegraphics[width=8.9cm]{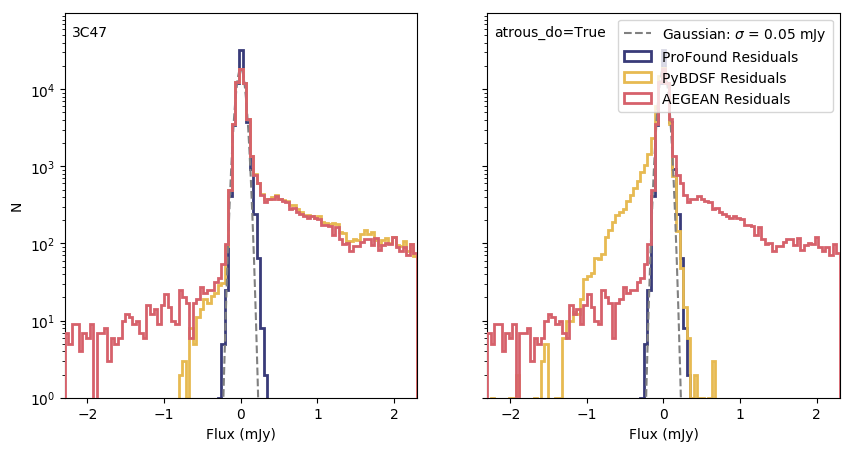}
\subcaption{3C47}
\end{minipage}%
\caption{Comparison of the residuals derived from the models from \textsc{ProFound} (blue), \textsc{PyBDSF} (yellow) and \textsc{AEGEAN} (red) of five 3C sources (whose images are shown in the left hand panel). Also shown is a Gaussian (fit for both amplitude and $\sigma$) used to give a modelled estimate of the noise. An estimate of $\sigma$ (in Jy), which is related to the noise is given in the legend in the top right hand corner. These are shown for the sources: 3C16 (a), 3C19 (b), 3C28 (c), 3C42 (d) and 3C47 (e) {as seen in Figure \ref{fig:3csources} (where the entire image for which these residuals are calculated over are shown)}. This is shown when the default parameters of \textsc{PyBDSF} and \textsc{AEGEAN} are used to generate the source models.}
\label{fig:3cresiduals}
\end{center}
\end{figure*}

To quantify how well \textsc{ProFound} is able to recover all the radio emission for these sources, again we investigate the residual image. If a source extractor has truly recovered the emission from the object, only noise should remain which should appear as a Gaussian distribution centred around zero. The results of this can be seen in Figure \ref{fig:3cresiduals}. Also shown is a model for Gaussian noise as a grey dashed line, this again used to highlight what typical Gaussian noise in the image should look like. This was modelled by fitting the histogram of the negative residuals from \textsc{ProFound}, fit for both amplitude and $\sigma$. 

{From Figure \ref{fig:3cresiduals} it can be seen that the residuals from \textsc{ProFound} are consistently well modelled as a Gaussian. This suggests it is successfully extracting the full fluxes of these sources. Although occasionally there are small excesses at high and low flux densities per beam (e.g. Figure \ref{fig:3cresiduals}(e)). There is also a peak in the histogram around a flux density per beam of 0 mJy/beam. This excess is again due to the smooth sky model that \textsc{ProFound} uses and so small noise fluctuations may be included as part of the source. This was not as obvious in the residuals from Section \ref{sec:blindcat} due to both the large number of pixels as well as the small covering factor of sources in the image. In these images of 3C sources, however, the source is a large fraction of the image and so this excess at 0 mJy/beam is obvious}. For \textsc{PyBDSF} and \textsc{AEGEAN} on the other hand, there are very clear excesses in the flux density per beam of the residuals at both high and low values. As the definition of residual is the image-model, at high flux densities per beam an excess represents where a source model has under predicted the flux density per beam whereas an excess at negative flux densities pre beam suggests that the Gaussian components have over-predicted the flux density per beam needed. With \texttt{atrous\_do} on for \textsc{PyBDSF}, it is able to model most of the emission (as there are typically few positive residuals) however there can be a large amount of negative residuals. This suggests that there is over-fitting of components where Gaussians are not as appropriate for the shape of the emission.

This work therefore highlights how \textsc{ProFound} is capable of tracing and modelling the emission from sources with known extended jet emission. It also highlights how using assumed Gaussian components may end up over-fitting such emission. 

\section{Discussions and Conclusions}
\label{sec:discussion}

In this paper we have shown the potential of using \textsc{ProFound} to detect and model the emission of sources from radio continuum images. \textsc{ProFound} was developed with the aim of optical and IR surveys, where noise is uncorrelated however we have shown in this paper that despite the heavily correlated noise in radio continuum imaging, pixel based extraction software are able to work well in this regime. Using \textsc{ProFound} does not assume a morphology, the flux of resolved sources can be better traced and as components of a certain morphology are not used, regions outside the source can not be over fit.

{\textsc{ProFound} has been tested in this paper through simulations of varying morphologies and consistently calculates accurate flux densities of sources. These morphologies were created using Gaussians, elliptical discs convolved with the beam and complex extended sources. Both \textsc{PyBDSF} and \textsc{AEGEAN} also succeeded well in recovering the flux densities of single objects (i.e. the Gaussian and elliptical sources) or smooth double lobed objects. However, they struggled in comparison to recover the flux densities of extended sources which have complex morphologies. }

By considering the residuals that remain in the images once sources have been removed, it is also evident that \textsc{ProFound} can successfully model the flux of sources. This was especially evident when five 3C sources were investigated where there was an excess of negative residuals for the other software. This is related to the fact that Gaussian components are not always appropriate to model these complex sources and may over fit the extended emission whilst also missing flux in other regions.

For current and future surveys there are both benefits to using source extractors that fit Gaussian components as well as pixel based source extraction. Fitting Gaussian components is especially useful for calibration purposes in building up sky models (an application we are not considering in this study). As well as this, for telescopes such as the MWA \citep{Tingay2013} as well as in single dish observations with e.g. Arecibo and the Green Bank Telescope (GBT), the resolution of these telescopes is constrained to arcminute resolution, and so images are likely to consist of unresolved sources which have a known shape given by the synthesised beam of the telescope. In this case where all the emission is typically unresolved, fitting Gaussian sources (of the beam shape) may seem as an appropriate method. Other radio facilities however such as the VLA, MeerKAT \citep{Jonas2016}, ASKAP \citep{Norris2011} and LOFAR \citep{LOFAR} resolve more structure to the AGN and SFGs they observe. In these cases \textsc{ProFound} models the full complexity of these sources, as shown in Figures \ref{fig:sources} and \ref{fig:3csources}. For surveys from these facilities \textsc{ProFound} may have an advantage by better modelling these complexities as well as combining multiple components of the same source together. This obviously will not work in cases where e.g. there are two lobed jets separated by a large separation, however these would not be merged together by any standard source finding algorithm. By also showing that \textsc{ProFound} successfully detects smooth Gaussian emission we suggest that \textsc{ProFound} is capable of accounting for and characterising the multitude of sources observed in radio surveys. 

We therefore feel that \textsc{ProFound} may be a beneficial source extraction software for both current as well as the future radio surveys that we expect to complete at higher angular resolutions and greater depths. Not only this, but as \textsc{ProFound} is designed to be used within a multi-wavelength framework. This can therefore be used to generate consistent flux extraction of sources across the electromagnetic spectrum. This is by using segments defined by \textsc{ProFound} at one wavelength to calculate fluxes at another. This will be useful for not only obtaining consistently extracted fluxes at different radio frequencies but can also be important in making use of observations across the electromagnetic spectrum. This is advantageous in the era of multi-wavelength astronomy. It also has the potential to use the ancillary information to make sub-threshold detections of radio sources, which we will discuss further in future work. 

\section*{Acknowledgements}
We thank the Referee for their useful comments with this work. CLH would like to acknowledge the support given from the Science and Technology Facilities Council (STFC) for their support to the first author through an STFC studentship {(ST/N504233/1)}. CLH is also grateful to STFC for their support and funding for a Long Term Attachment (LTA), which made this work possible. This work was also supported by the Oxford Hintze Centre for Astrophysical Surveys, which is funded through generous support from the Hintze Family Charitable Foundation and the award of the STFC consolidated grant (ST/N000919/1).



\pagebreak
\bibliographystyle{mnras}
\bibliography{radioprofound} 




\bsp	
\label{lastpage}
\end{document}